\documentclass[10pt,english]{article}
\usepackage{ae,aecompl}
\usepackage[T1]{fontenc}
\usepackage[latin9]{inputenc}
\usepackage{color}
\usepackage{babel}
\usepackage{bm}
\usepackage{amsmath}
\usepackage{amssymb}
\usepackage{graphicx}
\usepackage{esint}
\usepackage[unicode=true,pdfusetitle,
 bookmarks=true,bookmarksnumbered=false,bookmarksopen=false,
 breaklinks=false,pdfborder={0 0 1},backref=false,colorlinks=false]
 {hyperref}

\makeatletter

\let\SF@@footnote\footnote
\def\footnote{\ifx\protect\@typeset@protect
    \expandafter\SF@@footnote
  \else
    \expandafter\SF@gobble@opt
  \fi
}
\expandafter\def\csname SF@gobble@opt \endcsname{\@ifnextchar[
  \SF@gobble@twobracket
  \@gobble
}
\edef\SF@gobble@opt{\noexpand\protect
  \expandafter\noexpand\csname SF@gobble@opt \endcsname}
\def\SF@gobble@twobracket[#1]#2{}
\providecommand{\tabularnewline}{\\}

\usepackage{babel}
\usepackage[font=small,labelfont=bf]{caption}
\@ifundefined{definecolor}{\@ifundefined{definecolor}
 {\usepackage{color}}{}
}{}
\@ifundefined{definecolor}{\@ifundefined{definecolor}
 {\usepackage{color}}{}
}{}
\@ifundefined{definecolor}{\@ifundefined{definecolor}
 {\@ifundefined{definecolor}
 {\@ifundefined{definecolor}
 {\usepackage{color}}{}
}{}
}{}
}{}
\@ifundefined{definecolor}{\@ifundefined{definecolor}
 {\@ifundefined{definecolor}
 {\@ifundefined{definecolor}
 {\@ifundefined{definecolor}
 {\usepackage{color}}{}
}{}
}{}
}{}
}{}
\@ifundefined{definecolor}{\@ifundefined{definecolor}
 {\@ifundefined{definecolor}
 {\@ifundefined{definecolor}
 {\@ifundefined{definecolor}
 {\@ifundefined{definecolor}
 {\usepackage{color}}{}
}{}
}{}
}{}
}{}
}{}
\@ifundefined{definecolor}{\@ifundefined{definecolor}
 {\@ifundefined{definecolor}
 {\@ifundefined{definecolor}
 {\@ifundefined{definecolor}
 {\@ifundefined{definecolor}
 {\@ifundefined{definecolor}
 {\usepackage{color}}{}
}{}
}{}
}{}
}{}
}{}
}{}
\@ifundefined{definecolor}{\@ifundefined{definecolor}
 {\@ifundefined{definecolor}
 {\@ifundefined{definecolor}
 {\@ifundefined{definecolor}
 {\@ifundefined{definecolor}
 {\@ifundefined{definecolor}
 {\@ifundefined{definecolor}
 {\usepackage{color}}{}
}{}
}{}
}{}
}{}
}{}
}{}
}{}
\@ifundefined{definecolor}{\@ifundefined{definecolor}
 {\@ifundefined{definecolor}
 {\@ifundefined{definecolor}
 {\@ifundefined{definecolor}
 {\@ifundefined{definecolor}
 {\@ifundefined{definecolor}
 {\@ifundefined{definecolor}
 {\@ifundefined{definecolor}
 {\usepackage{color}}{}
}{}
}{}
}{}
}{}
}{}
}{}
}{}
}{}
\@ifundefined{definecolor}{\@ifundefined{definecolor}
 {\@ifundefined{definecolor}
 {\@ifundefined{definecolor}
 {\@ifundefined{definecolor}
 {\@ifundefined{definecolor}
 {\@ifundefined{definecolor}
 {\@ifundefined{definecolor}
 {\@ifundefined{definecolor}
 {\@ifundefined{definecolor}
 {\usepackage{color}}{}
}{}
}{}
}{}
}{}
}{}
}{}
}{}
}{}
}{}
\@ifundefined{definecolor}{\@ifundefined{definecolor}
 {\@ifundefined{definecolor}
 {\@ifundefined{definecolor}
 {\@ifundefined{definecolor}
 {\@ifundefined{definecolor}
 {\@ifundefined{definecolor}
 {\@ifundefined{definecolor}
 {\@ifundefined{definecolor}
 {\@ifundefined{definecolor}
 {\@ifundefined{definecolor}
 {\usepackage{color}}{}
}{}
}{}
}{}
}{}
}{}
}{}
}{}
}{}
}{}
}{}
\@ifundefined{definecolor}{\@ifundefined{definecolor}
 {\@ifundefined{definecolor}
 {\@ifundefined{definecolor}
 {\@ifundefined{definecolor}
 {\@ifundefined{definecolor}
 {\@ifundefined{definecolor}
 {\@ifundefined{definecolor}
 {\@ifundefined{definecolor}
 {\@ifundefined{definecolor}
 {\@ifundefined{definecolor}
 {\@ifundefined{definecolor}
 {\usepackage{color}}{}
}{}
}{}
}{}
}{}
}{}
}{}
}{}
}{}
}{}
}{}
}{}
\@ifundefined{definecolor}{\@ifundefined{definecolor}
 {\@ifundefined{definecolor}
 {\@ifundefined{definecolor}
 {\@ifundefined{definecolor}
 {\@ifundefined{definecolor}
 {\@ifundefined{definecolor}
 {\@ifundefined{definecolor}
 {\@ifundefined{definecolor}
 {\@ifundefined{definecolor}
 {\@ifundefined{definecolor}
 {\@ifundefined{definecolor}
 {\usepackage{color}}{}
}{}
}{}
}{}
}{}
}{}
}{}
}{}
}{}
}{}
}{}
}{}
\@ifundefined{definecolor}{\@ifundefined{definecolor}
 {\@ifundefined{definecolor}
 {\@ifundefined{definecolor}
 {\@ifundefined{definecolor}
 {\@ifundefined{definecolor}
 {\@ifundefined{definecolor}
 {\@ifundefined{definecolor}
 {\@ifundefined{definecolor}
 {\@ifundefined{definecolor}
 {\@ifundefined{definecolor}
 {\@ifundefined{definecolor}
 {\@ifundefined{definecolor}
 {\@ifundefined{definecolor}
 {\usepackage{color}}{}
}{}
}{}
}{}
}{}
}{}
}{}
}{}
}{}
}{}
}{}
}{}
}{}
}{}
\@ifundefined{definecolor}{\@ifundefined{definecolor}
 {\@ifundefined{definecolor}
 {\@ifundefined{definecolor}
 {\@ifundefined{definecolor}
 {\@ifundefined{definecolor}
 {\@ifundefined{definecolor}
 {\@ifundefined{definecolor}
 {\@ifundefined{definecolor}
 {\@ifundefined{definecolor}
 {\@ifundefined{definecolor}
 {\@ifundefined{definecolor}
 {\@ifundefined{definecolor}
 {\@ifundefined{definecolor}
 {\@ifundefined{definecolor}
 {\usepackage{color}}{}
}{}
}{}
}{}
}{}
}{}
}{}
}{}
}{}
}{}
}{}
}{}
}{}
}{}
}{}
\usepackage{bm}

\@ifundefined{definecolor}{\@ifundefined{definecolor}
 {\@ifundefined{definecolor}
 {\@ifundefined{definecolor}
 {\@ifundefined{definecolor}
 {\@ifundefined{definecolor}
 {\@ifundefined{definecolor}
 {\@ifundefined{definecolor}
 {\@ifundefined{definecolor}
 {\@ifundefined{definecolor}
 {\@ifundefined{definecolor}
 {\@ifundefined{definecolor}
 {\@ifundefined{definecolor}
 {\@ifundefined{definecolor}
 {\@ifundefined{definecolor}
 {\@ifundefined{definecolor}
 {\usepackage{color}}{}
}{}
}{}
}{}
}{}
}{}
}{}
}{}
}{}
}{}
}{}
}{}
}{}
}{}
}{}
}{}

\usepackage{bm}

\@ifundefined{definecolor}{\@ifundefined{definecolor}
 {\@ifundefined{definecolor}
 {\@ifundefined{definecolor}
 {\@ifundefined{definecolor}
 {\@ifundefined{definecolor}
 {\@ifundefined{definecolor}
 {\@ifundefined{definecolor}
 {\@ifundefined{definecolor}
 {\@ifundefined{definecolor}
 {\@ifundefined{definecolor}
 {\@ifundefined{definecolor}
 {\@ifundefined{definecolor}
 {\@ifundefined{definecolor}
 {\@ifundefined{definecolor}
 {\@ifundefined{definecolor}
 {\@ifundefined{definecolor}
 {\usepackage{color}}{}
}{}
}{}
}{}
}{}
}{}
}{}
}{}
}{}
}{}
}{}
}{}
}{}
}{}
}{}
}{}
}{}

\usepackage{amsfonts}

\usepackage{epsfig}

\usepackage{latexsym}

\def\lesssim{\mathrel{\hbox{\rlap{\hbox{\lower4pt\hbox{$\sim$}}}\hbox{$<$}}}}
\def\gtrsim{\mathrel{\hbox{\rlap{\hbox{\lower4pt\hbox{$\sim$}}}\hbox{$>$}}}}

\usepackage{aecompl}
\usepackage{cite}

\makeatother

\begin{document}
\title{\textbf{{Frame-dragging: meaning, myths, and misconceptions }}}
\author{L. Filipe O. Costa$^{1}$\thanks{lfilipecosta@tecnico.ulisboa.pt},
José Natário$^{1}$\thanks{jnatar@math.ist.utl.pt} \\
 \\
 \\
 {\em $^{1}$CAMGSD -- Departamento de Matemática, Instituto Superior
Técnico,} \\
 {\em Universidade de Lisboa, 1049-001, Lisboa, Portugal}\\
 }
\date{\today}

\maketitle
 
\begin{abstract}
Originally introduced in connection with general relativistic Coriolis
forces, the term \emph{frame-dragging} is associated today with a
plethora of effects related to the off-diagonal element of the metric
tensor. It is also frequently the subject of misconceptions leading
to incorrect predictions, even of nonexistent effects. We show that
there are three different levels of frame-dragging corresponding to
three distinct gravitomagnetic objects: gravitomagnetic potential
1-form, field, and tidal tensor, whose effects are independent, and
sometimes opposing. It is seen that, from the two analogies commonly
employed, the analogy with magnetism holds strong where it applies,
whereas the fluid-dragging analogy (albeit of some use, qualitatively,
in the first level) is, in general, misleading. Common misconceptions
(such as viscous-type ``body-dragging'') are debunked. Applications
considered include rotating cylinders (Lewis-Weyl metrics), Kerr,
Kerr-Newman and Kerr-dS spacetimes, black holes surrounded by disks/rings,
and binary systems. 
\end{abstract}
\tableofcontents{}

\section{Introduction}

The term ``dragging'' in the context of relativistic effects generated
by the motion of matter was first coined by Einstein in his 1913 letter
to Mach \cite{EinsteinLetterMach}, in connection with the general
relativistic Coriolis force generated in the interior of a spinning
mass shell, causing the plane of a Foucault pendulum to be ``dragged
around''. The term appeared again in the papers by Lense and Thirring
\cite{LenseThirringTranslated}, namely, a ``dragging coefficient''
was defined as the (inverse) ratio between the shell's angular velocity
($\Omega$), and the angular velocity ($\Omega'$) of the reference
frame for which the Coriolis forces vanish in its interior (corresponding
to the local inertial frame). Although the notion was already implicit
in these works, it was not, however, until Cohen's 1965 paper \cite{Cohen1965FrameDragging}
that the designation ``dragging of inertial frames'' first appeared;
therein $\Omega'$ was dubbed ``angular velocity of the inertial
frames'' inside the shell. The underlying principle is the same as
for the Coriolis forces that arise near a spinning body, in a reference
frame fixed to the distant stars. It is also the same for the precession
of gyroscopes placed therein \cite{Pugh1959,Schiff_PNAS1960}, in
which case one also talks about dragging of the ``compass of inertia''
(the compass of inertia being defined as a system of axes undergoing
Fermi-Walker transport, and physically realized precisely by the spin
vectors of a set of guiding gyroscopes \cite{MassaZordan,CiufoliniWheeler,GodelRotatingUniverses}).
All these effects can be assigned \cite{Costa:2015hlh} to the action
of a Coriolis or ``gravitomagnetic'' field generated by the source's
motion, and they are all commonly refereed to as ``frame-dragging''
effects (e.g. \cite{CiufoliniWheeler,Misner:1974qy,CiufoliniNature2007,ThorneDragging1971,BlackHolesTheMembrane,ThorneGravitomagnetism,Schaefer2004}).

Later, another class of effects \cite{BaardenPressTeukolsky,Misner:1974qy,BlackSaturn,CizekSemerakDisks2017}
started being dubbed frame-dragging as well. They pertain to axistationary
metrics, and include: the fact that the observers of zero angular
momentum (ZAMOs) have non-vanishing angular velocity in a coordinate
system fixed to the distant stars (dragging of the ZAMOs); and, conversely,
objects with zero angular velocity have non-zero angular momentum.
These are facets of a principle which, as we shall see, is different
from the frame-dragging class mentioned above.

More recently \cite{NicholsDragg}, yet another type of effect ---
a \emph{local} (i.e., tidal) one, stemming from the curvature tensor
--- has been rightfully dubbed frame-dragging: the precession of
a gyroscope with respect to a system of axes attached to guiding gyroscopes
at an infinitesimally close point (``differential precession'').

In order to gain intuition into the ``frame-dragging'' effects,
two analogies have been put forth: the electromagnetic analogy, in
particular, between the magnetic field and the general relativistic
Coriolis field (thus dubbed ``gravitomagnetic field''), and the
fluid-dragging analogy. The former is based on solid equations, best
known in weak field slow motion approximations \cite{CiufoliniWheeler,Damour:1990pi,Harris_1992,ohanian_ruffini_2013,Ruggiero:2002GMeffects,Gralla:2010xg,WillPoissonBook},
but with exact versions \cite{LandauLifshitz,ZonozBell1998,ManyFaces,MassaII,Cattaneo1958,NatarioQM2007,Analogies,Zonoz2019,Cilindros,BlackHolesTheMembrane,RizziRuggieroAharonovII}
holding in arbitrarily strong fields, and is known for providing a
familiar and reliable formalism. The fluid analogy, initially proposed
in \cite{Schiff_PNAS1960,SchiffPRL1960}, and then supported by other
authors (e.g. \cite{ThorneDragging1971,BlackHolesTheMembrane,CiufoliniWheeler}),
consists of an analogy drawn between the effects created by the rotation
of a body immersed in a fluid, and the frame-dragging effects. Albeit
providing (to a limited extent) a certain qualitative intuition for
the second class of effects mentioned above (dragging of the ZAMOs),
it is generically misleading and the source of most misconceptions
and incorrect predictions concerning frame-dragging. Some of these
were noticed already in the literature, most notably in a paper by
Rindler \cite{RINDLERDragging}, where, in the framework of a linearized
theory approximation, several inconsistencies of the fluid-dragging
model are pointed out, and the gravito-electromagnetic analogy is
recommended instead.

In this paper we start (Sec. \ref{sec:Distinct-effects-under}) by
observing that the three classes (or levels) of ``frame-dragging''
effects are governed by distinct mathematical objects (the gravitomagnetic
potential 1-form $\bm{\mathcal{A}}$, the gravitomagnetic field, and
the gravitomagnetic tidal tensor) corresponding to different orders
of differentiation of $\bm{\mathcal{A}}$, and underlying physical
principles (dragging of the ZAMOs, and dragging or differential dragging
of the compass of inertia); moreover, these levels are largely independent,
there existing solutions displaying only the first or second levels,
as well as phenomena where different levels of frame-dragging act
oppositely (Secs. \ref{subsec:Competing-effects} and \ref{subsec:Gravitomagnetic-tidal-effects:}).
Section \ref{sec:Frame-dragging-is-never} is devoted to debunking
the common wrong notion that when a source (e.g. a black hole) spins,
it forces test bodies around into rotation (``body-dragging'').
We start by generalizing Rindler's paper to the exact theory, using
the exact 1+3 ``gravitoelectromagnetic'' (GEM) formalism, exemplifying
with an imaginary space station around a spinning black hole. We consider
also the reciprocal problem (a rotating ring around a non-spinning
black hole), and, to clear any doubt that no such effect takes place,
the \emph{static} equilibrium positions for test particles in the
equatorial plane of spinning black hole spacetimes (Kerr-de Sitter
and Kerr-Newman). Finally, we consider a notable phenomenon driven
by frame-dragging --- bobbings in binary systems --- where the body-dragging
picture predicts the opposite of the true effect.

\emph{Notation and conventions.---} We use the signature $(-+++)$;\textcolor{black}{{}
Greek letters $\alpha$, $\beta$, $\gamma$, ... denote 4D spacetime
indices, running 0-3; Roman letters $i,j,k,...$ denote spatial indices,
running 1-3}; $\epsilon_{\alpha\beta\gamma\delta}\equiv\sqrt{-g}[\alpha\beta\gamma\delta]$
is the 4-D Levi-Civita tensor, with the orientation $[1230]=1$ (i.e.,
in flat spacetime, $\epsilon_{1230}=1$); $\epsilon_{ijk}\equiv\sqrt{h}[ijk]$\textcolor{black}{{}
is the} Levi-Civita tensor in a 3-D Riemannian manifold of metric
$h_{ij}$. Our convention for the Riemann tensor is $R_{\ \beta\mu\nu}^{\alpha}=\Gamma_{\beta\nu,\mu}^{\alpha}-\Gamma_{\beta\mu,\nu}^{\alpha}+...$
. $\star$ denotes the Hodge dual (e.g. $\star F_{\alpha\beta}\equiv\epsilon_{\alpha\beta}^{\ \ \ \mu\nu}F_{\mu\nu}/2$,
for a 2-form $F_{\alpha\beta}=F_{[\alpha\beta]}$). The basis vector
corresponding to a coordinate $\phi$ is denoted by $\partial_{\phi}\equiv\partial/\partial\phi$,
and its $\alpha$-component by $\partial_{\phi}^{\alpha}\equiv\delta_{\phi}^{\alpha}$.

\section{Distinct effects under the same denomination\label{sec:Distinct-effects-under}}

We will derive here the exact equations describing the different types
of frame-dragging (equations of ``gravitoelectromagnetism'', or
GEM), considering stationary spacetimes, where their formulation (the
so called 1+3 ``quasi-Maxwell'' formalism \cite{LandauLifshitz,ManyFaces,ZonozBell1998,NatarioQM2007,KatzBellBicakCQG2011,Analogies,Zonoz2019,Cilindros})
is particularly simple and intuitive. A generalization for arbitrary
time-dependent fields is given in Appendix \ref{Appendix:GEMgeneral}.

The line element $ds^{2}=g_{\alpha\beta}dx^{\alpha}dx^{\beta}$ of
a stationary spacetime can generically be written in the form 
\begin{equation}
ds^{2}=-e^{2\Phi}(dt-\mathcal{A}_{i}dx^{i})^{2}+h_{ij}dx^{i}dx^{j}\ ,\label{eq:StatMetric}
\end{equation}
where $e^{2\Phi}=-g_{00}$, $\Phi\equiv\Phi(x^{j})$, $\mathcal{A}_{i}\equiv\mathcal{A}_{i}(x^{j})=-g_{0i}/g_{00}$,
and $h_{ij}\equiv h_{ij}(x^{k})=g_{ij}+e^{2\Phi}\mathcal{A}_{i}\mathcal{A}_{j}$.
Observers of 4-velocity 
\begin{equation}
u^{\alpha}\equiv u_{{\rm lab}}^{\alpha}=(-g_{00})^{-1/2}\partial_{t}^{\alpha}=e^{-\Phi}\partial_{t}^{\alpha}\equiv e^{-\Phi}\delta_{0}^{\alpha}\label{eq:uLab}
\end{equation}
(i.e. whose worldlines are tangent to the timelike Killing vector
field $\partial_{t}$) are \emph{at rest} in the coordinate system
of (\ref{eq:StatMetric}); they shall be called ``laboratory'' observers.
The quotient of the spacetime by the worldlines of the laboratory
observers yields a 3-D manifold $\Sigma$ in which $h_{ij}$ is a
Riemannian metric, called the spatial or ``orthogonal'' metric.
It can be identified in spacetime with the projector orthogonal to
$u^{\alpha}$ (\emph{space projector} with respect to $u^{\alpha}$),
\begin{equation}
h_{\alpha\beta}\equiv u_{\alpha}u_{\beta}+g_{\alpha\beta}\ ,\label{eq:SpaceProjector}
\end{equation}
and yields the spatial distances between neighboring laboratory observers,
as measured through Einstein's light signaling procedure \cite{LandauLifshitz}.

\subsection{Sagnac effect and dragging of the ZAMOs \label{subsec:DraggingZAMOS}}

Unlike translational motion, which is inherently relative, rotation
is not, and is physically detectable. A way of detecting the \emph{absolute}
rotation of an apparatus (i.e., its rotation relative to the ``spacetime
geometry'' \cite{Misner:1974qy}, whose meaning shall be clear below)
is the Sagnac effect \cite{Post1967,Chow_et_al1985,AshtekarMagnon,Tartaglia:1998rh,Kajari:2009qy,Cilindros,RizziRuggieroAharonovII}.
It consists of the difference in arrival times of light-beams propagating
around a closed path in opposite directions. In flat spacetime, where
the concept was first introduced (see e.g. \cite{Post1967,Chow_et_al1985,Kajari:2009qy,RizziRuggieroAharonovII}
and references therein), the time difference is originated by the
rotation of the apparatus with respect to global inertial frames (thus
to the ``distant stars''), see e.g. Fig. 1 in \cite{Cilindros}.
In a gravitational field, however, it arises also in apparatuses which
are fixed relative to the distant stars (i.e., to \emph{asymptotically}
inertial frames) \cite{Chow_et_al1985,AshtekarMagnon,Tartaglia:1998rh,Kajari:2009qy,Cilindros};
in this case one talks about ``frame-dragging''. Both effects can
be read from the spacetime metric (\ref{eq:StatMetric}), encompassing
the flat Minkowski metric expressed in a rotating coordinate system,
as well as arbitrary stationary gravitational fields. Along a photon
worldline, $ds^{2}=0$; by (\ref{eq:StatMetric}), this yields two
solutions $dt=\mathcal{A}_{i}dx^{i}\pm e^{-\Phi}\sqrt{h_{ij}dx^{i}dx^{j}}$,
the $+$ sign corresponding to the future-oriented worldline;\footnote{The future-pointing condition is $k_{\alpha}\partial_{t}^{\alpha}=k_{0}<0\ \Leftrightarrow\ dt>\mathcal{A}_{i}dx^{i}$,
where $k^{\alpha}\equiv dx^{\alpha}/d\lambda$ is the vector tangent
to the photon's worldline.} therefore 
\[
dt=\mathcal{A}_{i}dx^{i}+e^{-\Phi}dl\ ,
\]
where $dl\equiv\sqrt{h_{ij}dx^{i}dx^{j}}$ is the spatial distance
element. Consider photons constrained to move within a closed loop
$C$ in the space manifold $\Sigma$; for instance, within an optical
fiber loop, as depicted in Fig. \ref{fig:ClockKerr}(a). Using the
$+$ ($-$) sign to denote the anti-clockwise (clockwise) directions,
the coordinate time it takes for a full loop is, respectively, $t_{\pm}=\oint_{\pm C}dt=\oint_{C}e^{-\Phi}dl\pm\oint_{C}\mathcal{A}_{i}dx^{i}$;
therefore, the Sagnac \emph{coordinate} time delay $\Delta t$ is
(e.g. \cite{Cilindros}) 
\begin{equation}
\Delta t_{{\rm S}}\equiv t_{+}-t_{-}=2\oint_{C}\mathcal{A}_{i}dx^{i}=2\oint_{C}\bm{\mathcal{A}}\ ,\label{eq:SagnacDiffForm}
\end{equation}
where we identified $\mathcal{A}_{i}dx^{i}$ with the 1-form $\bm{\mathcal{A}}\equiv\mathcal{A}_{i}\mathbf{d}x^{i}$
on the space manifold $\Sigma$.

Consider now an axistationary spacetime, whose line element \eqref{eq:StatMetric}
simplifies to (in spherical-type coordinates) 
\begin{equation}
ds^{2}=-e^{2\Phi}(dt-\mathcal{A}_{\phi}d\phi)^{2}+h_{rr}dr^{2}+h_{\theta\theta}d\theta^{2}+h_{\phi\phi}d\phi^{2}\ .\label{eq:AxistatMetric}
\end{equation}
In spite of being at rest, the laboratory observers \eqref{eq:uLab}
have, in general, non-zero angular momentum. The component of their
angular momentum along the symmetry axis is, per unit mass \cite{Misner:1974qy,SemerakGRG1998,Cilindros},
\begin{equation}
u_{\phi}=u^{0}g_{0\phi}=\frac{g_{0\phi}}{\sqrt{-g_{00}}}=e^{\Phi}\mathcal{A}_{\phi}\ ,\label{eq:AngMomentumLab}
\end{equation}
which is zero \emph{iff} $\mathcal{A}_{\phi}=0$. This manifests physically
as follows. Take an observer at $r=r_{0}$, and consider a circular
optical loop (e.g., an optical fiber) of the same radius around the
axis $\theta=0$, see Fig. \ref{fig:ClockKerr}(a). Such an observer
will measure a Sagnac effect; i.e., it will see light beams emitted
in opposite directions along the loop not completing it at the same
time, the difference in (coordinate) arrival times being, according
to Eq. (\ref{eq:SagnacDiffForm}), 
\begin{equation}
\Delta t_{{\rm S}}=2\oint_{C}\mathcal{A}_{\phi}d\phi=4\pi\mathcal{A}_{\phi}\ .\label{eq:DtBigloop}
\end{equation}
Only observers with zero angular momentum (ZAMOs) measure no Sagnac
effect. These observers are such that $(u_{{\rm ZAMO}})_{\phi}=0$,
i.e., have angular velocity 
\begin{equation}
\Omega_{{\rm ZAMO}}\equiv\Omega_{{\rm ZAMO}}(r,\theta)=\frac{u_{{\rm ZAMO}}^{\phi}}{u_{{\rm ZAMO}}^{0}}=-\frac{g_{0\phi}}{g_{\phi\phi}}\ .\label{eq:OmegaZamo}
\end{equation}
That the Sagnac effect vanishes for them can easily be seen by performing
a local coordinate transformation $\bar{\phi}=\phi-\Omega_{{\rm ZAMO}}(r_{0})t$,
leading to a coordinate system where the ZAMO at $r_{0}$ is at rest.
The metric form thereby obtained is diagonal \emph{at} $r_{0}$: $\bar{g}_{0\bar{\phi}}(r_{0})=0$,
hence $\Delta\bar{t}_{{\rm S}}=0$. This singles out the ZAMOs as
those who regard the $\pm\bar{\phi}$ directions as geometrically
equivalent; for this reason they are said to be those that do not
rotate with respect to ``the local spacetime geometry'' \cite{Misner:1974qy}.

If\footnote{Actually, the weaker condition that the congruence of observers at
rest \eqref{eq:uLab} is inertial at infinity suffices.} $g_{\alpha\beta}\stackrel{r\rightarrow\infty}{\rightarrow}\eta_{\alpha\beta}$,
the coordinate system in \eqref{eq:AxistatMetric} corresponds to
a rigid frame anchored to the asymptotic inertial frame at infinity
(i.e., to the distant stars). Hence, 
\begin{itemize}
\item the ``laboratory'' observers, at rest in a frame fixed to the distant
stars, have non-zero angular momentum \eqref{eq:AngMomentumLab} (measuring
a Sagnac effect); 
\item the zero angular momentum observers have non-zero angular velocity
\eqref{eq:OmegaZamo} in a coordinate system fixed to the distant
stars (or as ``viewed'' from an observer at infinity). 
\end{itemize}
These features are usually assigned to ``frame-dragging''; we point
out that it in fact consists of the dragging of the ZAMOs, and the
gravitomagnetic object governing it is the potential 1-form $\bm{\mathcal{A}}$
(or equivalently, the gravitomagnetic vector potential $\vec{\mathcal{A}}$).
These are the frame-dragging effects involved in arranging the bodies'
angular momentum/angular velocities in e.g. the black hole--ring
and black hole--disk systems in \cite{WillRing1974,CizekSemerakDisks2017},
as well as in black saturn systems \cite{BlackSaturn}, as discussed
in Sec. \ref{sec:Frame-dragging-is-never} and Fig. \ref{fig:Station}
below.

\subsection{Dragging of the compass of inertia: gravitomagnetic field and Lense-Thirring
effects\label{subsec:Dragging-of-theCompass}}

Consider a (point-like) test particle of worldline $x^{\alpha}(\tau)$,
4-velocity $dx^{\alpha}/d\tau\equiv U^{\alpha}$ and mass $m$. The
space components of the geodesic equation $DU^{\alpha}/d\tau=0$ yield\footnote{\label{fn:Christoffel}Computing the Christoffel symbols $\Gamma_{00}^{i}=-e^{2\Phi}G^{i}$,
$\Gamma_{j0}^{i}=e^{2\Phi}\mathcal{A}_{j}G^{i}-e^{\Phi}H_{\ j}^{i}/2$,
and $\Gamma_{jk}^{i}=\Gamma(h)_{jk}^{i}-e^{\Phi}\mathcal{A}_{(k}H_{j)}^{\ i}-e^{2\Phi}G^{i}\mathcal{A}_{j}\mathcal{A}_{k}$,
where $H_{ij}\equiv e^{\Phi}[\mathcal{A}_{j,i}-\mathcal{A}_{i,j}]$.}, for the line element \eqref{eq:StatMetric} \cite{LandauLifshitz,ZonozBell1998,NatarioQM2007,Analogies,Zonoz2019,Cilindros},
\begin{equation}
\frac{\tilde{D}\vec{U}}{d\tau}=\gamma\left[\gamma\vec{G}+\vec{U}\times\vec{H}\right]=\frac{\vec{F}_{{\rm GEM}}}{m}\ ,\label{eq:QMGeo}
\end{equation}
where $\gamma=-U^{\alpha}u_{\alpha}=e^{\Phi}(U^{0}-U^{i}\mathcal{A}_{i})$
is the Lorentz factor between $U^{\alpha}$ and $u^{\alpha}$, 
\begin{equation}
\left[\frac{\tilde{D}\vec{U}}{d\tau}\right]^{i}=\frac{dU^{i}}{d\tau}+\Gamma(h)_{jk}^{i}U^{j}U^{k}\ ;\qquad\Gamma(h)_{jk}^{i}=\frac{1}{2}h^{il}\left(h_{lj,k}+h_{lk,j}-h_{jk,l}\right)\label{eq:3DAccel}
\end{equation}
is the Levi-Civita covariant derivative with respect to the spatial
metric $h_{ij}$, with $\Gamma(h)_{jk}^{i}$ the corresponding Christoffel
symbols, and 
\begin{equation}
\vec{G}=-\tilde{\nabla}\Phi\ ;\qquad\quad\vec{H}=e^{\Phi}\tilde{\nabla}\times\vec{\mathcal{A}}\label{eq:GEMFieldsQM}
\end{equation}
are vector fields living on the space manifold $\Sigma$ with metric
$h_{ij}$, dubbed, respectively, ``gravitoelectric'' and ``gravitomagnetic''
fields. These play in Eq. (\ref{eq:QMGeo}) roles analogous to those
of the electric ($\vec{E}$) and magnetic ($\vec{B}$) fields in the
Lorentz force equation, $DU^{i}/d\tau=(q/m)[\gamma\vec{E}+\vec{U}\times\vec{B}]^{i}$.
The analogy motivates also dubbing $\vec{\mathcal{A}}$ \emph{gravitomagnetic
vector potential.} Here, $\tilde{\nabla}$ denotes covariant differentiation
with respect to the spatial metric $h_{ij}$ {[}i.e., the Levi-Civita
connection of $(\Sigma,h)${]}, Equation (\ref{eq:3DAccel}) is the
standard 3-D covariant acceleration, and Eq. (\ref{eq:QMGeo}) describes
the acceleration of the curve obtained by projecting the time-like
geodesic onto the space manifold $(\Sigma,h)$, with $\vec{U}$ as
its tangent vector {[}identified in spacetime with the projection
of $U^{\alpha}$ onto $(\Sigma,h)$: $(\vec{U})^{\alpha}=h_{\ \beta}^{\alpha}U^{\beta}$,
see Eq. (\ref{eq:SpaceProjector}){]}. The physical interpretation
of Eq. (\ref{eq:QMGeo}) is that, from the point of view of the laboratory
observers, the spatial trajectory will appear accelerated, as if acted
upon by the fictitious force $\vec{F}_{{\rm GEM}}$ (standing here
for ``gravitoelectromagnetic'' force\footnote{In \cite{Analogies} a different convention was used, in that $\vec{F}_{{\rm GEM}}$
(and the term ``inertial force'') therein actually refers to the
inertial force per unit mass (i.e., the inertial ``acceleration''
$\vec{F}_{{\rm GEM}}/m$, in the notation herein).}). In other words, the laboratory observers measure \emph{inertial}
\emph{forces}, which arise from the fact that the laboratory frame
is \emph{not inertial}; in fact, $\vec{G}$ and $\vec{H}$ are identified
in spacetime, respectively, with minus the acceleration and twice
the vorticity of the laboratory observers: 
\begin{equation}
G^{\alpha}=-\nabla_{\mathbf{u}}u^{\alpha}\equiv-u_{\ ;\beta}^{\alpha}u^{\beta}\;;\qquad H^{\alpha}=2\omega^{\alpha}=\epsilon^{\alpha\beta\gamma\delta}u_{\gamma;\beta}u_{\delta}\;.\label{eq:GEM Fields Cov}
\end{equation}
They may be regarded as the relativistic generalization of, respectively,
the Newtonian gravitational field and the classical Coriolis field,
encompassing them as limiting cases \cite{Costa:2015hlh}. It is $\vec{H}$
that governs, via Eq. \eqref{eq:QMGeo}, the Coriolis (i.e., gravitomagnetic)
forces generated inside a spinning hollow sphere, noted by Einstein
\cite{EinsteinLetterMach,PfisterHistory} and Thirring \cite{LenseThirringTranslated};
or those acting on test particles in the exterior field of a spinning
body, causing the Lense-Thirring orbital precession \cite{LenseThirringTranslated,CiufoliniWheeler}.
It governs also the ``precession'' of gyroscopes with respect to
the reference frame associated to the coordinate system in \eqref{eq:StatMetric}:
according to the Mathisson-Papapetrou equations \cite{Mathisson:1937zz,Papapetrou:1951pa,Dixon:1970zza,Gralla:2010xg,Costa:2012cy},
under the Mathisson-Pirani spin condition \cite{Mathisson:1937zz,Pirani:1956tn},
the spin vector $S^{\alpha}$ of a gyroscope (i.e., a spinning pole-dipole
particle) of 4-velocity $U^{\alpha}=dx^{\alpha}/d\tau$ is Fermi-Walker
transported along its center of mass worldline $x^{\alpha}(\tau)$,
\begin{equation}
\frac{D_{F}S^{\alpha}}{d\tau}=0\ \Leftrightarrow\ \frac{DS^{\alpha}}{d\tau}=S^{\mu}a_{\mu}U^{\alpha}\label{eq:GyrosFW}
\end{equation}
where $a^{\alpha}\equiv DU^{\alpha}/d\tau$. The spin vector is spatial
with respect to $U^{\alpha}$, $S^{\alpha}U_{\alpha}=0$, and so,
for a gyroscope whose center of mass is at rest in the coordinates
of (\ref{eq:StatMetric}), $U^{\alpha}=u^{\alpha}$ {[}see Eq. (\ref{eq:uLab}){]},
the space part of Eq. \eqref{eq:GyrosFW} reads (using the Christoffel
symbols in footnote \ref{fn:Christoffel}, and noting that $S^{\alpha}u_{\alpha}=0\Rightarrow S^{0}=S^{i}\mathcal{A}_{i}$)
\cite{NatarioQM2007,Analogies,ManyFaces,Cilindros} 
\begin{equation}
\frac{d\vec{S}}{d\tau}=\frac{1}{2}\vec{S}\times\vec{H}\ ,\label{eq:SpinPrec}
\end{equation}
resembling the precession of a magnetic dipole $\vec{\mu}$ in a magnetic
field, $D\vec{S}/d\tau=\vec{\mu}\times\vec{B}$. Likewise, the Sagnac
time delay in an optical gyroscope (i.e., a small optical loop $C$)
is also governed by the gravitomagnetic field $\vec{H}$, as can be
seen by applying the Stokes theorem to (\ref{eq:SagnacDiffForm}),
considering the surface $\mathcal{S}$ with boundary $\partial\mathcal{S}=C$,
\begin{equation}
\Delta t_{{\rm S}}=2\oint_{\partial\mathcal{S}}\bm{\mathcal{A}}=2\int_{\mathcal{S}}\mathbf{d}\bm{\mathcal{A}}=2\int_{\mathcal{S}}e^{-\Phi}H^{k}d\mathcal{S}_{k}\approx2e^{-\Phi}\vec{H}\cdot\vec{{\rm A}}\!{\rm rea}_{\mathcal{S}}\ ,\label{eq:SagnacStokes}
\end{equation}
where $\mathbf{d}\bm{\mathcal{A}}=\mathcal{A}_{j,i}\mathbf{d}x^{i}\wedge\mathbf{d}x^{j}=\epsilon_{ijk}H^{k}e^{-\Phi}\mathbf{d}x^{i}\wedge\mathbf{d}x^{j}/2$,
$d\mathcal{S}_{k}=\epsilon_{ijk}\mathbf{d}x^{i}\wedge\mathbf{d}x^{j}/2$
, $\epsilon_{ijk}=\sqrt{h}[ijk]$ is the Levi-Civita tensor of the
space manifold $(\Sigma,h)$, and $\vec{{\rm A}}\!{\rm rea}_{\mathcal{S}}$
the ``area vector'' of the loop (see \cite{Cilindros} and footnote
on p. 7 therein).

The gravito-electromagnetic analogy $\{\vec{G},\vec{H}\}\leftrightarrow\{\vec{E},\vec{B}\}$
also extends to the field equations: 
\begin{align}
 & \tilde{\nabla}\cdot\vec{G}=-4\pi(2\rho+T_{\ \alpha}^{\alpha})+{\vec{G}}^{2}+\frac{1}{2}{\vec{H}}^{2}\ ;\qquad\tilde{\nabla}\times\vec{G}=\ 0\ ;\label{eq:GFieldEq}\\
 & \tilde{\nabla}\cdot\vec{H}=-\vec{G}\cdot\vec{H}\;;\qquad\tilde{\nabla}\times\vec{H}=-16\pi\vec{J}+2\vec{G}\times\vec{H}\ ,\label{eq:HFieldEq}
\end{align}
where the equations for $\tilde{\nabla}\cdot\vec{G}$ and $\tilde{\nabla}\times\vec{H}$
are, respectively, the time-time and time-space projections of the
Einstein field equations $R_{\alpha\beta}=8\pi\left(T_{\alpha\beta}^{\ }-\frac{1}{2}g_{\alpha\beta}^{\ }T_{\ \gamma}^{\gamma}\right)$,
with $\rho\equiv T^{\alpha\beta}u_{\alpha}u_{\beta}$ and $J^{\alpha}\equiv-T^{\alpha\beta}u_{\beta}$,
and the equations for $\tilde{\nabla}\cdot\vec{H}$ and $\tilde{\nabla}\times\vec{G}$
follow directly from \eqref{eq:GEMFieldsQM}. They strongly resemble
the Maxwell equations in a rotating frame, see Table 2 of \cite{Analogies}.

Equation \eqref{eq:GyrosFW} tells us that the gyroscope's axis is
fixed with respect to a Fermi-Walker transported frame, which mathematically
defines a locally non-rotating frame (e.g. \cite{MassaZordan,Misner:1974qy});
it is said to follow the ``compass of inertia'' \cite{MassaZordan,CiufoliniWheeler,Analogies}.
This agrees with the notion that gyroscopes are objects that oppose
changes in the direction of their rotation axes. Hence, the gyroscope
``precession'' in \eqref{eq:SpinPrec} is thus in fact minus the
angular velocity of rotation $\vec{H}/2$ of the coordinate basis
vectors relative to a locally non-rotating frame. Consider now the
case that $g_{\alpha\beta}\stackrel{r\rightarrow\infty}{\rightarrow}\eta_{\alpha\beta}$,
in which the coordinate system in \eqref{eq:StatMetric} corresponds
to a rigid frame anchored to the asymptotic inertial frame at infinity.
So, at infinity, the reference frame is inertial; however, at finite
distance from the mass-energy currents that {[}by Eq. \eqref{eq:HFieldEq}{]}
source $\vec{H}$, one has, in general, $\vec{H}\ne0$, and so that
\emph{same} rigid frame is rotating (besides being accelerated, as
the observers at rest therein are not freely falling, $\nabla_{\mathbf{u}}u^{\alpha}=-G^{\alpha}\ne0$).
One can thus say that the motion of the sources (or mass-energy currents,
in general) \emph{drags} the local inertial frames, or the \emph{local
compass of inertia}. In some literature this is cast as the appearance
of vorticity \cite{Herrera_FrameDragging2006,Herrera_FrameDragging_and_SE2007},
the two notions being equivalent\footnote{Indeed, the vorticity $\omega^{\alpha}$ of a congruence of observers
corresponds precisely to the angular velocity of rotation of the connecting
vectors between neighboring observers with respect to axes Fermi-Walker
transported, see e.g. footnote in p. 7 of \cite{Cilindros}.} via Eq. \eqref{eq:GEM Fields Cov}.

\subsection{Competing effects --- circular geodesics\label{subsec:Competing-effects}}

Let $U^{\alpha}$ be the 4-velocity of a test particle moving along
an equatorial circular geodesic in an axistationary spacetime (with
reflection symmetry about the equatorial plane \cite{SajalReflection2021}),
and $\mathcal{L}=g_{\mu\nu}U^{\mu}U^{\nu}/2$ the corresponding Lagrangian.
The angular velocity $\Omega_{{\rm geo}}\equiv d\phi/dt=U^{\phi}/U^{0}$
of the circular geodesics is readily obtained from the Euler-Lagrange
equations, 
\begin{equation}
\frac{d}{d\tau}\left(\frac{\partial\mathcal{L}}{\partial U^{\alpha}}\right)-\frac{\partial\mathcal{L}}{\partial x^{\alpha}}=0\ ,\label{eq:EL}
\end{equation}
whose $r$-component $dU_{r}/d\tau=g_{\mu\nu,r}U^{\mu}U^{\nu}/2$
yields, for $U^{\alpha}=U^{0}(\delta_{0}^{\alpha}+\Omega_{{\rm geo}}\delta_{\phi}^{\alpha})$,
\[
g_{\phi\phi,r}\Omega_{{\rm geo}}^{2}+2g_{0\phi,r}\Omega_{{\rm geo}}+g_{00,r}=0\ .
\]
Its solution is 
\begin{equation}
\Omega_{{\rm geo}\pm}=\frac{-g_{0\phi,r}\pm\sqrt{g_{0\phi,r}^{2}-g_{\phi\phi,r}g_{00,r}}}{g_{\phi\phi,r}}\ ,\label{eq:OmegaGeo}
\end{equation}
the $+$ ($-$) sign corresponding, for $g_{\phi\phi,r}>0$ and $g_{00,r}<0$
(i.e., attractive $\vec{G}$), to prograde (retrograde) geodesics,
i.e., positive (negative) $\phi$ directions. This equation tells
us that, when $g_{0\phi}$ depends on $r$, the periods $t_{{\rm geo}\pm}=2\pi/|\Omega_{{\rm geo}\pm}|$
of prograde and retrograde geodesics differ; this effect has been
dubbed the gravitomagnetic ``clock effect'' \cite{COHENMashhoon1993,BonnorClockEffect,BiniJantzenMashhoon_Clock1,IorioClockEffect}.
The difference is given by 
\[
\Delta t_{{\rm geo}}=2\pi(\Omega_{{\rm geo+}}^{-1}+\Omega_{{\rm geo-}}^{-1})=-4\pi\frac{g_{0\phi,r}}{g_{00,r}}\ .
\]
Using $g_{0\phi}=-g_{00}\mathcal{A}_{\phi}$, noticing that reflection
symmetry implies, in the equatorial plane, $\mathcal{A}_{\phi,\theta}=0$,
we have, by \eqref{eq:GEMFieldsQM}, $\mathcal{A}_{\phi,r}=e^{-\Phi}\epsilon_{r\phi i}H^{i}=-e^{-2\Phi}\sqrt{-g}H^{\theta}$,
and so (cf. \cite{Cilindros}) 
\begin{figure}
\includegraphics[width=1\textwidth]{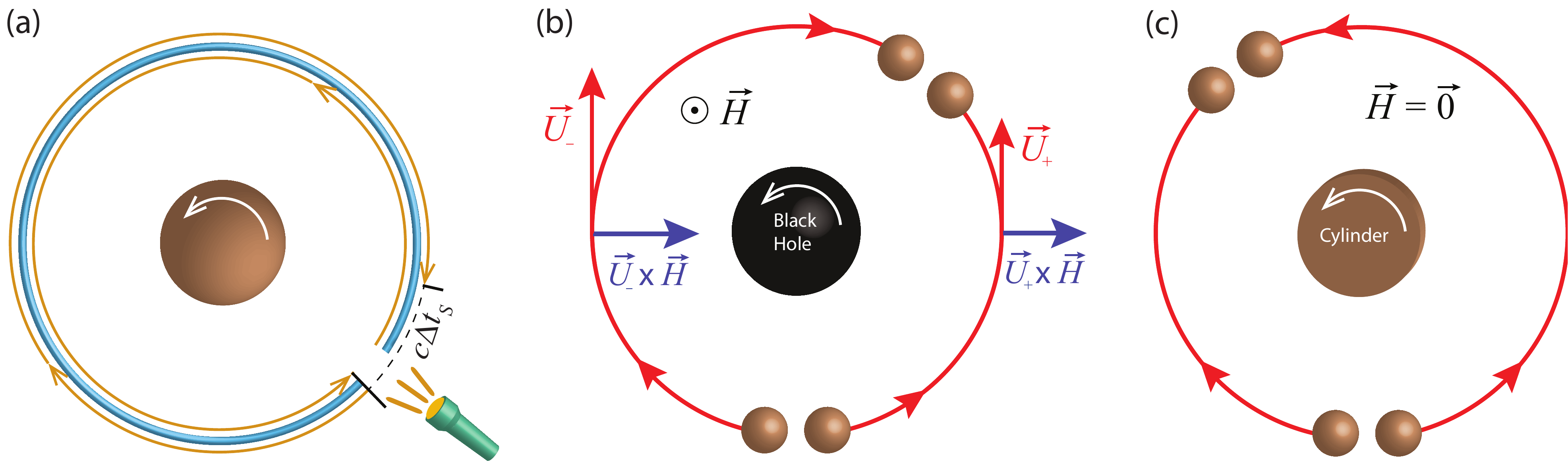} \caption{\label{fig:ClockKerr} (a) Sagnac effect in an optical loop around
a spinning body (dragging of the ZAMOs): a \textquotedblleft laboratory\textquotedblright{}
observer, at rest with respect to the distant stars, sends light beams
propagating in opposite directions along the loop; they take different
times to complete the loop, the co-rotating one arriving first, by
a time difference $\Delta t_{{\rm S}}<0$, Eq. (\ref{eq:DtBigloop}).
(b) For particles in circular geodesics around a Kerr black hole it
is the other way around: the co-rotating one has the longer period.
This is down to the combination of two oppositely competing effects:
the dragging of the ZAMOs, tending to decrease the period of the co-rotating
orbit vs. the gravitomagnetic force $m\gamma\vec{U}\times\vec{H}$
in Eq. (\ref{eq:QMGeo}) (dragging of the compass of inertia), which
is repulsive/attractive for co/counter-rotating orbits, thereby slowing/speeding
up the orbit, respectively. The latter effect prevails, so that $\Delta t_{{\rm geo}}=\Delta t_{{\rm S}}+\Delta t_{H}>0$.
(c) Around an infinite spinning cylinder of the Weyl class, $\vec{H}=0$;
hence, only the dragging of the ZAMOs subsists, and the situation
for circular geodesics is opposite to (b) {[}thus similar to (a){]}:
the co-rotating geodesic has the shortest period, the difference reducing
exactly to the Sagnac time delay, $\Delta t_{{\rm geo}}=\Delta t_{{\rm S}}<0$.}
\end{figure}

\begin{equation}
\Delta t_{{\rm geo}}=\Delta t_{{\rm S}}+\Delta t_{H};\qquad\Delta t_{{\rm S}}=4\pi\mathcal{A}_{\phi};\qquad\Delta t_{H}=\frac{2\pi\sqrt{-g}}{G_{r}e^{2\Phi}}H^{\theta}\ .\label{eq:GMClock}
\end{equation}
Hence, the gravitomagnetic clock effect consists of the sum of two
contributions originating from the two types of frame-dragging in
Secs. \ref{subsec:DraggingZAMOS} and \ref{subsec:Dragging-of-theCompass}:
the Sagnac time delay (\ref{eq:DtBigloop}) around the circular loop,
due to the dragging of the ZAMOs, governed by $\mathcal{A}_{\phi}$,
plus a term due to the gravitomagnetic (or Coriolis) forces generated
by the dragging of the compass of inertia, governed by the gravitomagnetic
field $\vec{H}$. The physical interpretation of the latter is as
follows: for circular orbits, the gravitomagnetic force $m\gamma\vec{U}\times\vec{H}$
in Eq. (\ref{eq:QMGeo}) is radial (since $\vec{H}=H^{\theta}\partial_{\theta}$
and $\vec{U}=U^{\phi}\partial_{\phi}$), being centrifugal or centripetal
depending on the $\pm\phi$ direction of the orbit, thus respectively
decreasing or increasing the overall attraction, and, consequently,
the velocity of the orbits. See Figure \ref{fig:ClockKerr}(b).

It is important to notice that the two contributions in \eqref{eq:GMClock}
are independent. In fact, there are solutions for which $\vec{H}$
vanishes whilst $\vec{\mathcal{A}}$ is non-zero, as is the case of
the Lewis metric of the Weyl class, describing the exterior gravitational
field produced by infinitely long rotating cylinders. The metric is
given, in star-fixed (``canonical'') coordinates, by Eq. (61) of
\cite{Cilindros}, yielding, in the form \eqref{eq:StatMetric}, 
\[
e^{2\Phi}=\frac{r^{4\lambda_{{\rm m}}}}{\alpha};\quad\bm{\mathcal{A}}=-\frac{j}{1/4-\lambda_{{\rm m}}}{\bf d}\phi;\quad h_{rr}=h_{zz}=r^{4\lambda_{{\rm m}}(2\lambda_{{\rm m}}-1)};\quad h_{\phi\phi}=\alpha r^{2(1-2\lambda_{{\rm m}})}\ ,
\]
$h_{ik}=0$ for $i\ne k$. Here $0\le\lambda_{{\rm m}}<1/4$ and $j$
are, respectively, the Komar mass and angular momentum per unit length,
and $\alpha$ the parameter governing the angle deficit ($j>0$, for
a cylinder spinning in the positive $\phi$ direction). Trivially
$\mathbf{d}\bm{\mathcal{A}}=0\Rightarrow\vec{H}=0$, hence $\Delta t_{H}=0$
and so 
\[
\Delta t_{{\rm geo}}=\Delta t_{{\rm S}}=4\pi\mathcal{A}_{\phi}=-\frac{4\pi j}{1/4-\lambda_{{\rm m}}}<0\ ,
\]
i.e., the difference in period between co- and counter-rotating circular
geodesics equals precisely the Sagnac time delay for photons, the
co-rotating one having a shorter period.

The two contributions can even be opposing, as is the case for the
Kerr metric. We have in this case, in the equatorial plane, 
\begin{equation}
e^{2\Phi}=1-\frac{2M}{r};\quad G_{r}=-\frac{M}{r^{2}-2Mr};\quad g=-r^{4};\quad\mathcal{A}_{\phi}=\frac{2aM}{2M-r};\quad\vec{H}=-\frac{2aM}{r^{3}(r-2M)}\partial_{\theta};\label{eq:QMKerr}
\end{equation}
and so (observe that always\footnote{The innermost circular geodesics, in each direction, are the photon
orbits whose radius is $r_{{\rm ph}\pm}=2M\{1+\cos[2\arccos(\mp a/M)/3]\}$
\cite{BaardenPressTeukolsky}, and so $r_{{\rm ph}-}\ge3M$.} $r>2M$, in order for the counter-rotating orbit to exist) 
\begin{equation}
\Delta t_{{\rm geo}}=\Delta t_{{\rm S}}+\Delta t_{H}=4\pi a\ ;\qquad\Delta t_{{\rm S}}=-\frac{8\pi aM}{r-2M}\,(<0)\qquad\Delta t_{H}=\frac{4\pi ar}{r-2M}\,(>0)\ ,\label{eq:DtKerr}
\end{equation}
i.e., the two contributions have \emph{opposite signs}. Namely, the
dragging of the ZAMOs tends to decrease the period of the orbit co-rotating
with the black hole by $\Delta t_{{\rm S}}$, as compared to the counter-rotating
orbit; however, the gravitomagnetic force $m\gamma\vec{U}\times\vec{H}$
(dragging the compass of inertia) does the opposite, tending to increase
the period of the co-rotating orbit (case in which $\vec{U}\times\vec{H}$
is repulsive), as compared to the counter-rotating orbit (case in
which $\vec{U}\times\vec{H}$ is attractive), by $\Delta t_{H}$,
see Fig. \ref{fig:ClockKerr}(b). Since $\Delta t_{H}>-\Delta t_{{\rm S}}$,
it is the gravitomagnetic force that prevails, making the co-rotating
orbits slower overall than the counter-rotating ones.

The gravitomagnetic force and the corresponding $\Delta t_{H}$ have
a close electromagnetic analogue in the magnetic force $q\vec{U}\times\vec{B}$
exerted on charged test particles orbiting a spinning charged body,
see Eq. (30) of \cite{Cilindros} --- only with a different sign,
which manifests that (anti-) parallel mass/energy currents have a
repulsive (attractive) gravitomagnetic interaction, as opposed to
magnetism, where (anti-) parallel charge currents attract (repel)
(see \cite{schutz_2003} and Sec. 13.6 of \cite{Feynman:1963uxa}).

\subsection{Gravitomagnetic \textquotedblleft tidal\textquotedblright{} effects:
\textquotedblleft differential\textquotedblright{} dragging and force
on gyroscopes\label{subsec:Gravitomagnetic-tidal-effects:}}

A third class of frame-dragging effects that has been (more recently)
discussed in the literature \cite{NicholsDragg,RuggieroGMRessonance,ThorneBlandfordBook},
distinct from those in Secs. \ref{subsec:DraggingZAMOS} and \ref{subsec:Dragging-of-theCompass},
is the ``differential precession'' of gyroscopes. It consists of
the precession of a gyroscope relative to a frame attached to the
spin axes of guiding gyroscopes at a neighboring point. The effect
was originally derived\footnote{In a perhaps less straightforward manner though, and with unnecessary
restrictions. Namely, in \cite{NicholsDragg} it is assumed that (besides
being momentary comoving) the gyroscopes at $L$ and $X_{2}^{i}$
have the same acceleration. This is not necessary, as shown here;
in order for (\ref{eq:RelPrec}) to hold, one needs only $U_{2}^{i}=0$,
i.e., that gyroscope 2 has momentarily zero ``Fermi relative velocity''
\cite{BolosIntrinsic} with respect to gyroscopes at $L$. Moreover,
the results therein hold only for vacuum, as the magnetic part of
the Weyl tensor is used instead of $\mathbb{H}_{\alpha\beta}$.} in \cite{NicholsDragg}; we briefly re-derive it below in a straightforward
manner, using Fermi coordinates.

The spin vector of a gyroscope in a gravitational field is Fermi-Walker
transported along its worldline, according to the Mathisson-Papapetrou-Pirani
Equation \eqref{eq:GyrosFW}. Consider an orthonormal tetrad ${\bf e}_{\hat{\alpha}}$
Fermi-Walker transported along the worldline $L(\tau)$ of the set
of gyroscopes 1. There is a coordinate system $\{X^{\alpha}\}$, rectangular
at $L$, and adapted to such tetrad ($\partial/\partial X^{\alpha}|_{L}={\bf e}_{\hat{\alpha}}$),
the so-called \cite{SyngeGenRel1960} ``Fermi coordinates'', where
the metric takes, to order $O(X^{2})$, locally the form 
\begin{equation}
ds^{2}=-\left[(1+a_{i}X^{i})^{2}+R_{0i0j}X^{i}X^{j}\right]dT^{2}-\frac{4}{3}R_{0jik}X^{j}X^{i}dTdX^{j}+\left(\delta_{ij}-\frac{1}{3}R_{iljm}X^{l}X^{m}\right)dX^{i}dX^{j}\label{eq:FermiMetric}
\end{equation}
(cf. e.g. Eq. (18) in \cite{NiZimmermannFermi1978}, setting therein
$\omega^{i}=0$). Here $a^{\alpha}=DU^{\alpha}/d\tau$ is the acceleration
of the fiducial worldline $L(\tau)=L(T)$, and $R_{\alpha\beta\gamma\delta}\equiv R_{\alpha\beta\gamma\delta}(T)$
are components of the curvature tensor evaluated along $L$. Let $\mathbf{e}_{\alpha}\equiv\partial/\partial X^{\alpha}$
denote the coordinate basis vectors and $\Gamma_{\beta\gamma}^{\alpha}$
its Christoffel symbols, $\Gamma_{\beta\gamma}^{\alpha}\mathbf{e}_{\alpha}=\nabla_{\mathbf{e}_{\beta}}\mathbf{e}_{\gamma}$.
Along $L$, the vectors $\mathbf{e}_{\alpha}$ are Fermi-Walker transported,
so $\left\langle \nabla_{\mathbf{e}_{0}}\mathbf{e}_{i},\mathbf{e}_{j}\right\rangle |_{X^{k}=0}=\Gamma_{0i}^{j}(L)=0$.
Hence gyroscopes moving along $L$ (by definition), or momentarily
at rest ($U^{i}=0$) at some event $\mathcal{P}_{1}:(T_{1},X_{1}^{i})=(T_{1},0)$
of $L$, do not precess relative to this frame, $d\vec{S}/dT|_{X^{i}=0}=D_{F}\vec{S}/d\tau|_{X^{i}=0}=0$,
by virtue of Eq. \eqref{eq:GyrosFW}. However, at some location $X_{2}^{i}$
outside $L$ ($X_{2}^{i}\ne0$), we have $\Gamma_{0i}^{j}(X_{2})=R_{\ ik0}^{j}X_{2}^{k}\ne0$,
and so the basis vectors $\mathbf{e}_{\alpha}$ are no longer Fermi-Walker
transported. That means that gyroscope 2, at the location $X_{2}^{i}$,
will precess with respect to this coordinate system. If the gyroscope
is therein at rest ($U_{2}^{i}=0$), we have 
\begin{equation}
\frac{dS_{2}^{i}}{d\tau_{2}}=-\Gamma_{0j}^{i}(X_{2})S_{2}^{j}U_{2}^{0}=-R_{\ jk0}^{i}X_{2}^{k}S_{2}^{j}=-R_{\ j\gamma\tau}^{i}U^{\tau}\delta x^{\gamma}S_{2}^{j}\ ,\label{eq:RelPrecGrav00}
\end{equation}
where in the first equality we noticed that $S_{2}^{0}=O(X^{2})$
and, in the second, that $U_{2}^{0}=(-g_{00})^{-1/2}=1+O(X^{2})$,
while only terms to first order in $X$ are to be kept in (\ref{eq:RelPrecGrav00})
to the accuracy at hand. In the last equality we noticed that $U^{\alpha}=\delta_{0}^{\alpha}$
and that, by the definition of the coordinate system $\{X^{\alpha}\}$
(see e.g. Fig. 13.4 of \cite{Misner:1974qy}), $X_{2}^{k}=\delta x^{k}$\textbf{
}are components of the vector $\delta x^{\alpha}=(0,\delta x^{i})$
at $L$ tangent to the (unique) spatial geodesic emanating orthogonally
from $L$ and passing through the event $\mathcal{P}_{2}:(T_{2},X_{2}^{i})$,
whose length equals that of the geodesic segment. It can be interpreted
as the ``separation vector'' between $\mathcal{P}_{2}$ and the
simultaneous ($T_{2}=T_{1}$) event $\mathcal{P}_{1}\in L$. Using
the space projection relation (cf. Eq. (5) in \cite{Analogies}) $h_{\alpha}^{\mu}h_{\beta}^{\nu}R_{\mu\nu\gamma\tau}=\epsilon_{\mu\alpha\beta\lambda}U^{\lambda}\star\!R_{\ \nu\gamma\tau}^{\mu}U^{\nu}$
which, in the coordinate system $\{X^{\alpha}\}$ (orthonormal at
$L$), reads: $R_{ij\gamma\tau\ }=\epsilon_{ij\mu\lambda}\star\!R_{\ \nu\gamma\tau}^{\mu}U^{\nu}U^{\lambda}$,
we have 
\begin{equation}
\frac{dS_{2}^{i}}{d\tau_{2}}=-\epsilon_{\ jk0}^{i}\delta\Omega^{k}S_{2}^{j}\quad\Leftrightarrow\quad\frac{d\vec{S}_{2}}{dT}=\delta\vec{\Omega}\times\vec{S}_{2}\ ,\qquad\delta\Omega^{k}\equiv\mathbb{H}_{\ \gamma}^{k}\delta x^{\gamma}\ ,\label{eq:RelPrec}
\end{equation}
where $\mathbb{H}_{\alpha\beta}\equiv\star R_{\alpha\mu\beta\nu}U^{\mu}U^{\nu}=\epsilon_{\alpha\mu}^{\ \ \sigma\lambda}R_{\sigma\lambda\beta\nu}U^{\mu}U^{\nu}/2$
is the ``gravitomagnetic tidal tensor'' (or ``magnetic'' part
of the Riemann tensor) as measured along $L$, and in the second equation
we noted that $d\vec{S}_{2}/d\tau_{2}=d\vec{S}_{2}/dT+O(X^{3})$.
Thus, $\delta\vec{\Omega}_{{\rm G}}$ is the angular velocity of precession
of gyroscopes at location $X_{2}^{i}$ with respect to the Fermi frame
locked to the guiding gyroscopes at the neighboring worldline $L$
(of course, this is just \emph{minus} the angular velocity of rotation
of the basis vectors $\mathbf{e}_{i}|_{X_{2}}$ relative to Fermi-Walker
transport). We can cast this effect as a \emph{differential dragging}
of the compass of inertia.

Another effect governed by the gravitomagnetic tidal tensor is the
spin-curvature force exerted on a gyroscope, described by the Mathisson-Papapetrou
equation \cite{Mathisson:1937zz,Papapetrou:1951pa,Dixon:1970zza,Gralla:2010xg,Costa:2012cy}
\begin{equation}
F^{\alpha}\equiv\frac{DP^{\alpha}}{d\tau}=-\frac{1}{2}R_{\ \beta\mu\nu}^{\alpha}S^{\mu\nu}U^{\beta}=-\mathbb{H}^{\beta\alpha}S_{\beta}\ ,\label{eq:SpinCurvature}
\end{equation}
where $S^{\alpha\beta}$ is the body's spin tensor, and in the second
equality we employed the Mathisson-Pirani \cite{Mathisson:1937zz,Pirani:1956tn}
spin condition $S^{\alpha\beta}U_{\beta}=0$, under which one can
write $S^{\mu\nu}=\epsilon^{\mu\nu\tau\lambda}S_{\tau}U_{\lambda}$.
This force is of different nature from the inertial GEM ``forces''
in the geodesic Equation \eqref{eq:QMGeo}, in that it is a physical,
covariant force, causing the body's 4-momentum $P^{\alpha}$ to change,
and its motion to be non-geodesic, $DU^{\alpha}/d\tau\ne0$.

Both Eqs. \eqref{eq:RelPrec} and \eqref{eq:SpinCurvature} have exact
(up to constant factors) electromagnetic analogues in terms of the
magnetic tidal tensor $B_{\alpha\beta}=\star F_{\alpha\mu;\beta}U^{\mu}$
\cite{Costa:2012cy,CostaHerdeiro}, namely, the differential precession
of magnetic dipoles $\delta\Omega_{{\rm EM}}^{i}=-\sigma B_{\ \gamma}^{i}\delta x^{\gamma}$
($\sigma\equiv\mu/S$) \cite{AnalogiesPreprint3}, and the force on
a magnetic dipole $F_{{\rm EM}}^{\alpha}=B^{\beta\alpha}\mu_{\beta}$
\cite{Costa:2012cy,CostaHerdeiro,Analogies}.

Another manifestation is in the geodesic deviation equation $D^{2}\delta x^{\alpha}/d\tau^{2}=-R_{\ \beta\gamma\delta}^{\alpha}\delta x^{\gamma}U^{\beta}U^{\delta}$
(e.g. \cite{CiufoliniWheeler,Misner:1974qy}). In vacuum, one can
decompose the Riemann tensor in terms of the gravitoelectric $(\mathbb{E}^{u})_{\alpha\beta}\equiv R_{\alpha\mu\beta\nu}u^{\mu}u^{\nu}$
and gravitomagnetic $(\mathbb{H}^{u})_{\alpha\beta}=\star R_{\alpha\mu\beta\nu}u^{\mu}u^{\nu}$
tidal tensors as measured by some laboratory observer $u^{\alpha}$,
see decomposition (44) of \cite{Costa:2012cy}; hence, with respect
to such an observer, the relative acceleration of nearby test particles
(of 4-velocity $U^{\alpha}$) comes in part from $(\mathbb{H}^{u})_{\alpha\beta}$,
which could thus be measured through gravity gradiometers \cite{MashhoonPaikWill}.

Finally, observe that, around the origin of the Fermi coordinate system
in \eqref{eq:FermiMetric}, the gravitomagnetic field as given by
Eq. \eqref{eq:GEM Fields Cov}, $H^{\alpha}=\epsilon_{\ \ \gamma\delta}^{\alpha\beta}u_{\ ;\beta}^{\gamma}u^{\delta}$,
$u^{\alpha}=(-g_{00})^{-1/2}\delta_{0}^{\alpha}$ (which is generally
valid, cf. Appendix \ref{Appendix:GEMgeneral}), reads 
\[
H^{i}=\epsilon_{\ \ k0}^{ij}u_{\ ;j}^{k}=\epsilon_{\ \ k0}^{ij}\Gamma_{0j}^{k}=\epsilon_{\ \ k0}^{ij}R_{\ jl0}^{k}X^{l}=-2\mathbb{H}_{\ l}^{i}X^{l}\quad\Leftrightarrow\quad\mathbb{H}_{\ j}^{i}=-\frac{1}{2}\partial_{j}H^{i}\ .
\]
Moreover, in the coordinate system of the stationary metric \eqref{eq:StatMetric},
we have \cite{Analogies} 
\begin{equation}
\mathbb{H}_{ij}=-\frac{1}{2}\left[\tilde{\nabla}_{j}H_{i}+(\vec{G}\cdot\vec{H})h_{ij}-2G_{j}H_{i}\right]\label{eq:HijGEM}
\end{equation}
(for a $\mathbb{H}_{\alpha\beta}$---$\vec{H}$ relation valid for
orthonormal frames in arbitrary spacetimes, see Eq. (110) of \cite{Analogies}).
We can thus say that $\mathbb{H}_{\alpha\beta}$ is essentially {[}to
leading order, in the case of \eqref{eq:HijGEM}{]} a derivative of
the gravitomagnetic field $\vec{H}$. So, we can cast the different
frame-dragging effects in the literature into the three levels of
gravitomagnetism in Table \ref{tab:Levels}, corresponding to three
different orders of differentiation of the gravitomagnetic 1-form
$\bm{\mathcal{A}}$. 
\begin{table*}
\begin{tabular}{|c|l|}
\hline 
\multicolumn{2}{|c|}{\raisebox{3.6ex}{}\raisebox{0.7ex}{\textbf{Levels of Gravitomagnetism/``Frame-Dragging''}}}\tabularnewline
\hline 
\hline 
\raisebox{3ex}{}\raisebox{0.5ex}{Governing gravitomagnetic object}  & \raisebox{0.5ex}{Physical effect}\tabularnewline
\hline 
\raisebox{2.5ex}{%
\begin{tabular}{c}
$\vec{\mathcal{A}}$\tabularnewline
{\footnotesize{}{}{}(gravitomagnetic}\tabularnewline
\raisebox{0.5ex}{{\footnotesize{}{}{}vector potential)}}\tabularnewline
\end{tabular}}  & \raisebox{3ex}{%
\begin{tabular}{l}
Dragging of the ZAMOs:\tabularnewline
$\bullet$ Sagnac effect\tabularnewline
\raisebox{2.5ex}{}$\bullet$ part of gravitomagnetic\tabularnewline
\raisebox{0.5ex}{~~~ ``clock effect''}\tabularnewline
\end{tabular}}\tabularnewline
\hline 
\raisebox{3.5ex}{%
\begin{tabular}{c}
\raisebox{0.1ex}{$\vec{H}$}\tabularnewline
{\footnotesize{}{}{}(gravitomagnetic}\tabularnewline
{\footnotesize{}{}{}field $=e^{\phi}\nabla\times\vec{\mathcal{A}}$)}\tabularnewline
\end{tabular}}  & \raisebox{2.5ex}{%
\begin{tabular}{l}
Dragging of the compass of inertia:\tabularnewline
$\bullet$ gravitomagnetic force\tabularnewline
\raisebox{0.5ex}{~~~$m\gamma\vec{U}\times\vec{H}$}\tabularnewline
$\bullet$ gyroscope precession\tabularnewline
\raisebox{0.5ex}{~~~$d\vec{S}/d\tau=\vec{S}\times\vec{H}/2$}\tabularnewline
$\bullet$ local Sagnac effect in\tabularnewline
\raisebox{0.5ex}{~~~light gyroscope}\tabularnewline
$\bullet$ part of gravitomagnetic\tabularnewline
\raisebox{0.5ex}{~~~ ``clock effect''}\tabularnewline
\end{tabular}}\tabularnewline
\hline 
\raisebox{2.5ex}{%
\begin{tabular}{c}
$\mathbb{H}_{\alpha\beta}$\tabularnewline
{\footnotesize{}{}{}(gravitomagnetic}\tabularnewline
\raisebox{0.5ex}{{\footnotesize{}{}{}tidal tensor $\sim\partial_{i}\partial_{j}\mathcal{A}_{k}$)}}\tabularnewline
\end{tabular}}  & \raisebox{2.5ex}{%
\begin{tabular}{l}
$\bullet$ Differential precession of gyroscopes\tabularnewline
\raisebox{0.5ex}{}~~~{\small{}{}{}${\displaystyle \delta\Omega^{i}=\mathbb{H}_{\ \beta}^{i}\delta x^{\beta}}$}\tabularnewline
\raisebox{2.5ex}{}$\bullet$ force on gyroscope\tabularnewline
\raisebox{0,5ex}{~~~{\small{}{}{}${\displaystyle DP^{\alpha}/d\tau=-\mathbb{H}^{\beta\alpha}S_{\beta}}$}}\tabularnewline
\end{tabular}}\tabularnewline
\hline 
\end{tabular}\caption{\label{tab:Levels}The different effects under the denomination \textquotedblleft frame-dragging\textquotedblright ,
cast into three levels of gravitomagnetism, corresponding to different
orders of differentiation of the gravitomagnetic vector potential
$\vec{\mathcal{A}}$.}
\end{table*}

We emphasize that these levels are independent: there exist solutions
where only the first level ($\bm{\mathcal{A}}$) is intrinsically
non-zero, as we have seen in Sec. \ref{subsec:Competing-effects};
and others possessing the first two levels, but where the third vanishes.
Examples of the latter are the Gödel universe and the uniform Som-Raychaudhuri
metrics which, as discussed in detail in Sec. VII.B.3 of \cite{Invariants},
are stationary solutions possessing zero gravitoelectric field ($\vec{G}=0$)
and non-zero \emph{uniform} gravitomagnetic field $\vec{H}$, leading,
by virtue of Eq. \eqref{eq:HijGEM}, $\mathbb{H}_{\alpha\beta}$ to
vanish for the rest observers therein, whilst both $\vec{H}\ne0$
and $\bm{\mathcal{A}}\ne0$. In such metrics, by Eq. \eqref{eq:SpinCurvature},
no force acts on gyroscopes at rest (whose worldlines are geodesic);
and, by Eq. \eqref{eq:RelPrec}, they moreover do not precess with
respect to neighboring guiding gyroscopes.

\section[Frame-dragging is never \textquotedblleft draggy\textquotedblright{}
--- no body-dragging]{Frame-dragging is never \textquotedblleft draggy\textquotedblright\protect\protect\footnote{Inspired on the title of the session \protect\protect\href{https://youtu.be/ho31IgLNxu8}{PT5}---
\textquotedblleft Dragging is never draggy: MAss and CHarge flows
in GR\textquotedblright{} (where \textquotedblleft draggy\textquotedblright{}
had however no such meaning), held at the sixteenth Marcel Grossmann
Meeting (MG16), July 5-10 2021.} --- no body-dragging\label{sec:Frame-dragging-is-never}}

It is a widespread myth that when a massive body (e.g. a black hole)
rotates, it drags everything around it. It seemingly originates from
the fluid dragging analogy proposed originally in \cite{Schiff_PNAS1960,SchiffPRL1960}
and reinforced in \cite{ThorneDragging1971,BlackHolesTheMembrane},
disseminating in the subsequent literature that the dragging of inertial
frames can be explained in analogy with the dragging of a viscous
fluid by an immersed rotating body. It is sometimes also portrayed
as the whirlpool analogy \cite{BlackHolesTheMembrane,RaineThomasBH}.
Such mental pictures can be extremely misleading \cite{RINDLERDragging,RaineThomasBH,ohanian_ruffini_2013};
frame-dragging, in none of its levels, produces that type of effects.
What is dragged are the ZAMOs, and the compass of inertia. The former
leads to Sagnac and other akin global effects (having no local effect
on the motion of test particles); the latter originates Coriolis (or
gravitomagnetic) inertial forces on test particles. Such forces, however,
affect only bodies \emph{moving} with respect to the chosen reference
frame, and are \emph{orthogonal} to the body's velocity; hence, \emph{never}
of the viscous/dragging type. This later aspect was stressed in an
illuminating paper by Rindler \cite{RINDLERDragging}, based on linearized
theory, in the framework of a weak field and slow motion approximation.

\begin{figure}[t]
\includegraphics[width=1\textwidth]{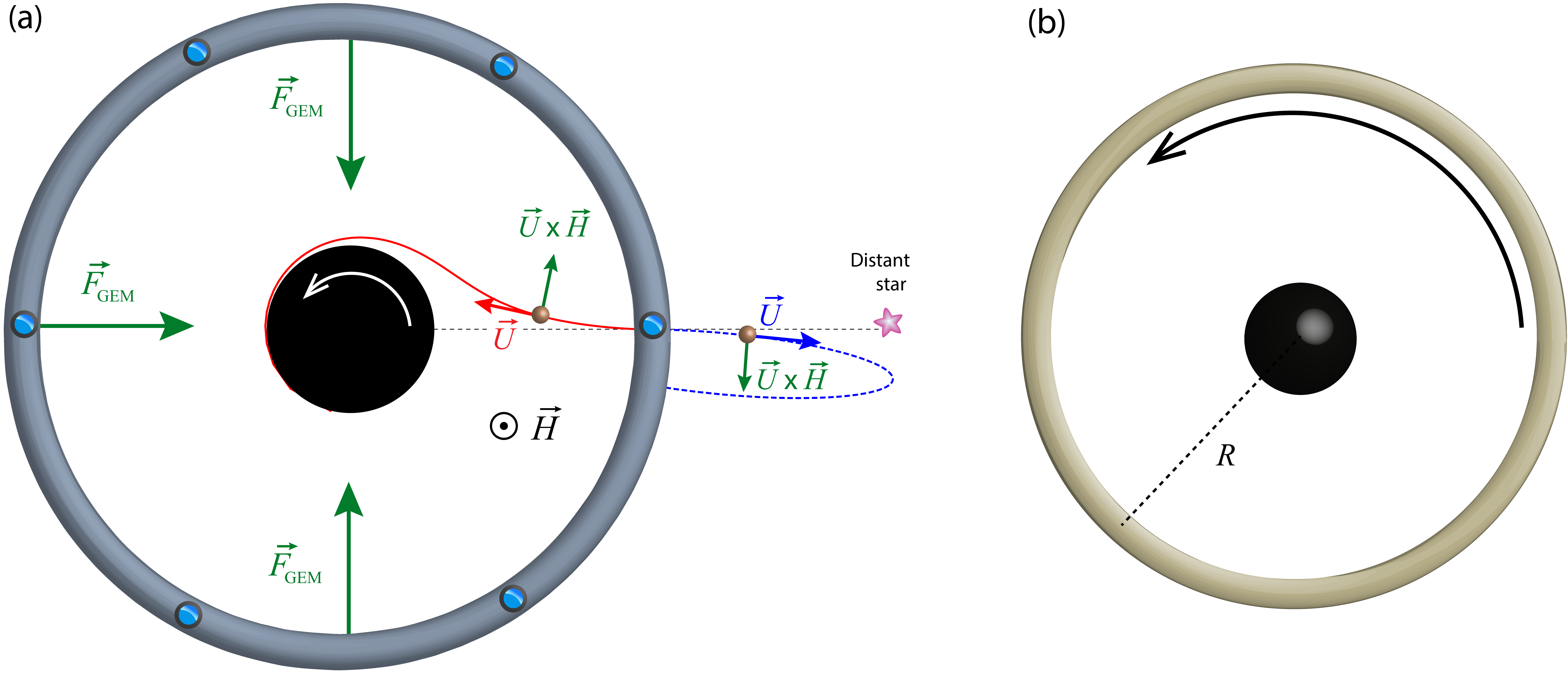} \caption{\label{fig:Station}(a) A space station (radius $R=5M)$ around a
spinning Kerr black hole ($a=0.9M$). It is not \textquotedblleft dragged\textquotedblright{}
around: if initially static, it remains so, each hatch pointing to
the same distant star. The dragging of the compass of inertia does
not affect it: the gravitomagnetic (i.e., Coriolis) forces vanish,
the gravitational forces (\ref{eq:QMGeo}) exerted on the station
reducing to gravitoelectric (i.e. \textquotedblleft Newtonian\textquotedblright )
forces, $\vec{F}_{{\rm GEM}}/m=\gamma^{2}\vec{G}$, which are \emph{radial}.
It is affected only by the dragging of the ZAMOs, causing it to have
a non-zero angular momentum, manifesting e.g. in a Sagnac effect along
optical fiber loops around the station. A particle dropped from rest
is deflected, along its infall, in the direction of the BH's rotation,
by the gravitomagnetic force $m\gamma\vec{U}\times\vec{H}$; however,
a particle launched with initial outwards radial velocity ($v=0.3$
for the blue dashed trajectory), is deflected in the \emph{opposite}
direction, contradicting again the naive body-dragging picture. {[}Red
and blue dashed trajectories are plots in Boyer-Lindquist coordinates
obtained by numerically solving the geodesic equation{]}. (b) A rotating
ring around a `non-spinning' BH. The BH does not acquire any angular
acceleration, and the configuration remains stationary. The consequence
of the ring's rotation is (via the dragging of the ZAMOs) causing
a zero angular momentum BH ($J=0)$; to have non-zero horizon angular
velocity ($\Omega_{{\rm BH}}\protect\ne0$) and, conversely, a BH
with zero angular velocity to have non-zero angular momentum.}
\end{figure}

Rindler's conclusion can be readily generalized to the exact theory
using Eq. \eqref{eq:QMGeo} {[}for stationary spacetimes, or with
Eq. \eqref{eq:GEMforce} of Appendix \ref{Appendix:GEMgeneral} for
arbitrary fields{]}. Imagine that an advanced civilization builds
a circular space station around a Kerr black hole, as illustrated
in Fig. \ref{fig:Station} (a), and imagine that it is initially set
fixed with respect to the distant stars (which can be set up by pointing
a telescope to a distant star). Regardless of the black hole's rotation,
the station will not be dragged into any rotational motion: since,
initially, $\vec{U}=0$, by Eq. \eqref{eq:QMGeo} the only inertial
force (per unit mass) acting on any of its mass elements is the radial
force produced by the gravitoelectric field (i.e., by the relativistic
generalization of the Newtonian field): $\vec{F}_{{\rm GEM}}/m=\gamma^{2}\vec{G}=\gamma^{2}G^{r}\partial_{r}$;
this force is counteracted by the stresses in the station's structure
(which prevent it from collapsing), and so $\vec{U}=0$ for all mass
elements at all times. That is, the station remains static, with e.g.
its hatches pointing to the same fixed stars. It is, in particular,
unaffected by the dragging of the compass of inertia. The situation
is only distinct from that around a static black hole in that here
the station's angular momentum per unit mass in non-zero {[}cf. Eq.
\eqref{eq:AngMomentumLab}{]}, 
\[
u_{\phi}|_{R}=(e^{\Phi}\mathcal{A}_{\phi})|_{R}=-\frac{2aM}{R\sqrt{1-2M/R}}\qquad(R\equiv\text{ station's radius})
\]
implying, e.g., that a Sagnac effect will be measured in an optical
fiber loop along the station, see Sec. \ref{subsec:DraggingZAMOS}
and Fig. \ref{fig:ClockKerr}(a) {[}the difference in arrival times
for beams propagating in opposite directions along the station being
given by $\Delta t_{{\rm S}}$ in \eqref{eq:DtKerr}, making therein
$r=R${]}. It implies also that it is impossible for the crew members
along the station to all have their clocks synchronized (see e.g.
\cite{Cilindros}, Secs. 5.3.2-5.3.3).

Imagine now that a crew member starts throwing objects (test particles);
by having a non-zero velocity ($\vec{U}\ne0$) with respect to the
star-fixed reference frame, gravitomagnetic (i.e., Coriolis) forces
$m\gamma\vec{U}\times\vec{H}$ will now act on them, cf. Eq. \eqref{eq:QMGeo}.
The gravitomagnetic field $\vec{H}$ in the equatorial plane of the
Kerr spacetime is given by Eq. \eqref{eq:QMKerr}; it is orthogonal
to the plane, pointing up. Consider, in particular, as depicted in
Fig. \ref{fig:Station} (a), one test particle dropped from rest from
the station, and another one launched with an initial outwards radial
velocity. The in-falling particle is indeed deflected in the direction
of the black hole's rotation; however the outgoing one is deflected
in the \emph{opposite} direction, contradicting the naive fluid dragging
picture. This exemplifies how different Coriolis forces are from viscous
dragging-type forces. In other words, the dragging of the compass
of inertia is a phenomenon very different from ``body-dragging.''

Another example of this difference is provided by the circular geodesics
around the black hole, depicted in Fig. \ref{fig:ClockKerr} (b):
the orbits are \emph{stationary}, their motion is not accelerated/decelerated
by the black hole's rotation (as would be the case if there were some
viscous coupling). The Coriolis forces $m\gamma\vec{U}\times\vec{H}$
are in this case \emph{radial}, consequently just changing the overall
gravitational attraction. Since they are repulsive (attractive) for
co-(counter-) rotating orbits, their effect is to actually (for a
fixed radius $r$) make the co-rotating geodesic \emph{slower} than
the counter-rotating one, somewhat at odds with the naive dragging
picture, as pointed out in \cite{RaineThomasBH,ohanian_ruffini_2013,IorioPhenomenologyLT}.

It is also instructive to consider the reciprocal of the setting in
Fig. \ref{fig:Station} (a), i.e., a non-spinning black-hole perturbed
by a rotating ring or disk around it, as depicted in Fig. \ref{fig:Station}
(b). The gravitational field produced by such setting is given by
the perturbative treatments in \cite{WillRing1974} (for a ring),
and in \cite{CizekSemerakDisks2017} (for a disk perturbing a Schwarzschild
black hole). Again, regardless of the ring/disk's rotating motion,
the black hole is not dragged around, in the sense of acquiring any
``angular acceleration''. In fact, these solutions are \emph{stationary}.
The impact of ring/disk's rotation is (via the dragging of the ZAMOs)
causing a black hole with zero angular momentum to have a non-zero
horizon angular velocity \cite{CizekSemerakDisks2017,WillRing1974,Hod:2015aua}
and, conversely, a non-rotating black hole to have a non-zero angular
momentum \cite{WillRing1974,Hod:2015aua} --- thereby even introducing
a certain ambiguity in a black hole's spinning/non-spinning character.
Let us see this in detail. First observe that $(u_{{\rm ZAMO}})_{\alpha}\propto\delta_{\alpha}^{0}=\nabla_{\alpha}t$
, and so the ZAMOs are orthogonal to the hypersurfaces of constant
time $t$, i.e., such hypersurfaces are their rest spaces. We can
thus say that, at each point, these hypersurfaces are rotating, with
respect to Boyer-Lindquist coordinates (anchored to inertial observers
at infinity) with an angular velocity $\Omega_{{\rm ZAMO}}(r,\theta)=-g_{0\phi}/g_{\phi\phi}$,
cf. Eq. \eqref{eq:OmegaZamo} (also called the ``dragging angular
velocity'' e.g. \cite{CizekSemerakDisks2017}). The horizon is a
2-surface embedded in a hypersurface of constant $t$, having the
special property that therein $\Omega_{{\rm ZAMO}}(r_{+})\equiv\Omega_{{\rm BH}}$
becomes independent\footnote{In any stationary black hole spacetime whose matter content obeys
the weak energy condition and hyperbolic equations, the horizon is
orthogonal to a (null) Killing vector field, i.e., it is a ``Killing
horizon'' (see \cite{HawkingBH1971}, \cite{NatarioMathRel} theorem
2.2).} of $\theta$; so, the horizon rotates rigidly with angular velocity
$\Omega_{{\rm BH}}$, sometimes dubbed the ``black hole's angular
velocity'' \cite{Wald:1984,Misner:1974qy,NatarioMathRel}. The black
hole's angular momentum is given \cite{WillRing1974,BardeenBook1973,CizekSemerakDisks2017}
by the Komar integral associated with the axisymmetry Killing vector
field $\xi^{\alpha}=\partial_{\phi}^{\alpha}$: $J=-(1/16\pi)\int_{\mathcal{S}_{{\rm BH}}}\star\mathbf{d}\bm{\xi}$,
for $\mathcal{S}_{{\rm BH}}$ some 2-surface enclosing the black hole
but \emph{not} the ring (in the case of the disk, $\mathcal{S}$ is
the BH horizon). The ring's angular momentum is $J_{\text{R}}=-(1/16\pi)\int_{\mathcal{S}}\star\mathbf{d}\bm{\xi}-J$,
for $\mathcal{S}$ enclosing the whole system. For a BH-ring system,
when the BH's angular momentum is zero ($J=0$), the horizon angular
velocity is non-zero, and given by Eqs. (75) and (68) of \cite{WillRing1974},
\[
\Omega_{{\rm BH}}|_{J=0}=\frac{2J_{{\rm R}}}{R^{3}}\ ,
\]
where $R$ is the ring's circumferential radius. Conversely, when
the horizon's angular velocity is zero ($\Omega_{{\rm BH}}=0$), the
BH's angular momentum is non-zero (negative): 
\[
J|_{\Omega_{{\rm BH}}=0}=-\frac{\pi^{-3/2}A_{{\rm BH}}^{3/2}}{8R^{3}}J_{{\rm R}}\ ,
\]
where $A_{{\rm BH}}$ is the horizon's area, cf. Eqs. (76), (68),
and (26) of \cite{WillRing1974}.

It is also worth mentioning that entirely analogous conclusions to
the above can be drawn using ``black saturns'' \cite{BlackSaturn},
which are exact (4+1)-dimensional solutions describing a black hole
surrounded by a black ring.

\subsection{Test particles in \emph{static} equilibrium around spinning black
holes}

Perhaps the sharpest counter-examples to the notion of body-dragging,
and how drastically actual frame-dragging (in any of its levels) differs
from it, is the existence of static equilibrium positions for test
particles around some spinning black hole spacetimes, such as the
Kerr-de Sitter and (for charged particles) the Kerr-Newman spacetimes.
The metrics of both these spacetimes are encompassed in the line element
(Kerr-Newman-dS metric \cite{VeselyZofka2019}) 
\begin{align}
 & ds^{2}=-\frac{\Delta_{r}}{\chi^{2}\Sigma}\left(dt-a\sin^{2}\theta d\phi\right)^{2}+\frac{\Sigma}{\Delta_{r}}dr^{2}+\frac{\Sigma}{\Delta_{\theta}}d\theta^{2}+\frac{\Delta_{\theta}\sin^{2}\theta}{\chi^{2}\Sigma}\left[adt-(a^{2}+r^{2})d\phi\right]^{2};\label{eq:KerrNewmandS}\\
 & {\displaystyle \Delta_{r}\equiv r^{2}-2Mr+a^{2}+Q^{2}-\frac{\Lambda}{3}r^{2}(r^{2}+a^{2})}\ ;\qquad{\displaystyle \chi\equiv1+\frac{\Lambda}{3}a^{2}}\ ;\label{KN-dSfunctions1}\\
 & {\displaystyle \Delta_{\theta}=1+\frac{\Lambda}{3}a^{2}\cos^{2}\theta}\ ;\qquad\Sigma\equiv r^{2}+a^{2}\cos^{2}\theta\ ,\label{KN-dSfunctions2}
\end{align}
where $\Lambda$ is the cosmological constant, and $Q$ the black
hole's charge.

\subsubsection{Kerr-de Sitter spacetime}

This metric is obtained by setting $Q=0$ in \eqref{eq:KerrNewmandS}-\eqref{KN-dSfunctions2}.
The gravitoelectric field {[}i.e., minus the acceleration of the laboratory
observers, Eq. (\ref{eq:GEM Fields Cov}){]} is, in the equatorial
plane ($\theta=\pi/2$), 
\[
\vec{G}=-\frac{(3M-r^{3}\Lambda)\Delta_{r}}{3r^{2}(\Delta_{r}-a^{2})}\partial_{r}\ .
\]
It vanishes at $r=\sqrt[3]{3M\Lambda^{-1}}\equiv r_{{\rm static}}$
(``static radius'' \cite{Stuchlik2004KerrdS}), where the cosmological
repulsion exactly balances the gravitational attraction. By virtue
of Eq. (\ref{eq:QMGeo}), this means that no gravitational inertial
force is exerted on a particle at rest at $r=r_{{\rm static}}$, $\vec{F}_{{\rm GEM}}=m\tilde{D}\vec{U}/d\tau=0$;
i.e. particles placed there remain static ($dx^{\alpha}/d\tau=0$),
while following a \emph{geodesic worldline} ($DU^{\alpha}/d\tau=d^{2}x^{\alpha}/d\tau^{2}=0$).
They are not dragged in any way by the black hole's rotation, no tangential
force being needed to hold them in place, unlike one might think based
on the viscous fluid dragging analogy. Frame-dragging has no detectable
effect on such particles, only causing them (via the dragging of the
ZAMOs) to have non-zero angular momentum, 
\[
U_{\phi}=u_{\phi}=-\frac{a(r^{2}+a^{2}-\Delta_{r})}{r\chi\sqrt{\Delta_{r}-a^{2}}}\ .
\]

\subsubsection{Kerr-Newman spacetime}

An analogous situation occurs in the Kerr-Newman spacetime, in this
case for charged particles. This metric is obtained by setting $\Lambda=0$
in \eqref{eq:KerrNewmandS}-\eqref{KN-dSfunctions2}; the corresponding
electromagnetic 4-potential 1-form is $\mathbf{A}=Qr(a\sin^{2}\theta\mathbf{d}\phi-\mathbf{d}t)/\Sigma$.
Consider now a (point-like) test particle of 4-velocity $U^{\alpha}$,
mass $m$, and charge $q$. Its equation of motion is $DU^{\alpha}d\tau=(q/m)F^{\alpha\beta}U_{\beta}$,
whose space part, in the ``laboratory frame'', reads, cf. Eqs. \eqref{eq:QMGeo}
and \eqref{eq:NonGeo}, 
\[
\frac{\tilde{D}\vec{U}}{d\tau}=\gamma\left[\gamma\vec{G}+\vec{U}\times\vec{H}\right]+\frac{q}{m}\left[\gamma\vec{E}+\vec{U}\times\vec{B}\right]\ ,
\]
where $\vec{E}$ and $\vec{B}$ denote, respectively, the space components
of the electric field $(E^{u})^{\alpha}\equiv F^{\alpha\beta}u_{\beta}$
and magnetic field $(B^{u})^{\alpha}\equiv\star F^{\alpha\beta}u_{\beta}$
as measured by the laboratory observers \eqref{eq:uLab}. In the equatorial
plane, $\vec{E}$ and $\vec{G}$ read 
\[
\vec{E}=\frac{Q\Delta_{r}}{r^{3}\sqrt{\Delta_{r}-a^{2}}}\partial_{r}\ ;\qquad\vec{G}=\frac{(Q^{2}-Mr)\Delta_{r}}{r^{3}(\Delta_{r}-a^{2})}\ .
\]
The equilibrium positions are given by the condition $\tilde{D}\vec{U}/d\tau=0=\vec{U}$;
noticing that $\vec{U}=0\Rightarrow U^{\alpha}=u^{\alpha}\Rightarrow\gamma=1$
{[}cf. Eq. \eqref{eq:uLab}, and recall that $\gamma\equiv-U^{\alpha}u_{\alpha}${]},
this condition becomes (in the equatorial plane) 
\[
\vec{G}=-\frac{q}{m}\vec{E}\quad\Leftrightarrow\quad\frac{qQ}{m}\sqrt{\Delta_{r}-a^{2}}=Mr-Q^{2}\ ,
\]
which, in the non-naked scenario $M^{2}>Q^{2}+a^{2}$, and outside
the ergosphere, $r>M+\sqrt{M^{2}-Q^{2}}$, yields the single solution
\cite{Aguirregabiria_1996} 
\[
r_{{\rm static}}=\frac{Q^{2}}{M-|q|\sqrt{(M^{2}-Q^{2})/(q^{2}-m^{2})}}\ ,
\]
with $Q$ and $q$ having the same sign (as expected), and $|q|>m$.
Hence, a particle with such a charge, placed at $r=r_{{\rm static}}$,
remains at rest, with no gravitomagnetic nor magnetic force acting
on it since $\vec{U}=0\Rightarrow\vec{U}\times\vec{H}=\vec{U}\times\vec{B}=0$.

\subsection{\textquotedblleft Bobbings\textquotedblright{} in binary systems\label{subsec:Bobbings-in-binary}}

\begin{figure}
\includegraphics[width=1\textwidth]{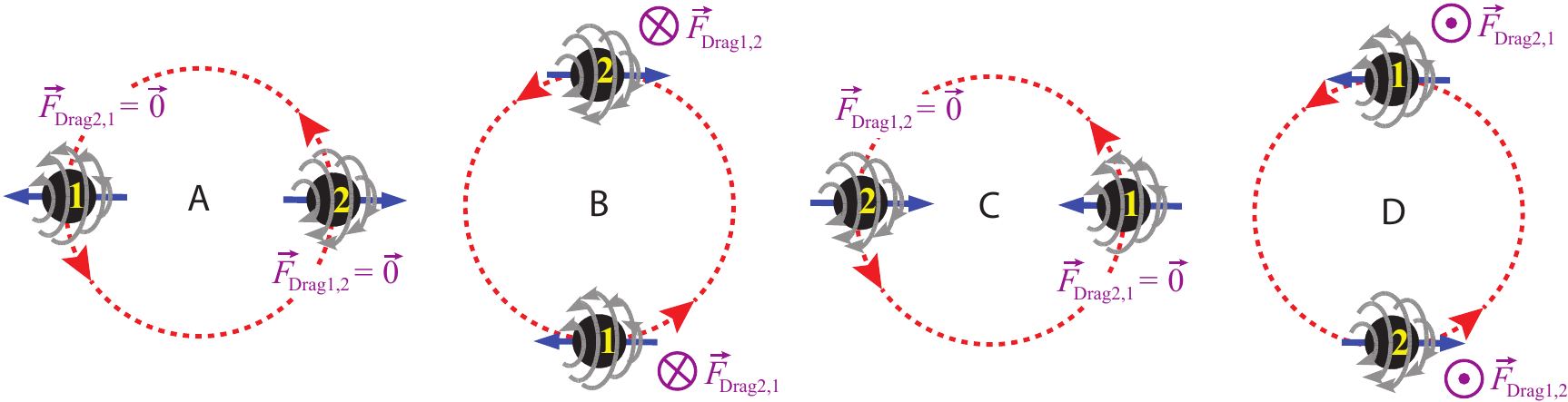} \caption{\label{fig:FalseBobbing}Incorrect prediction of the bobbing motions
of \textquotedblleft extreme kick\textquotedblright{} spinning black
hole binaries, as would follow from the body-dragging misconception
(illustration based on Fig. 5 of \cite{Pretorius2009}). Each black
hole would be acted by a viscous-type dragging force ($\vec{F}_{{\rm Drag}}$)
originated by the spin of the other black hole; such forces would
vanish when the black holes' spins lie along the axis connecting them
(phases A and C), and be maximum when the spins are orthogonal to
such an axis (phases B and D), each black hole dragging the other
inwards the plane of the illustration in phase B, and outwards in
phase D. \emph{None of these forces exist}, however.}
\end{figure}

\begin{figure}
\includegraphics[width=1\textwidth]{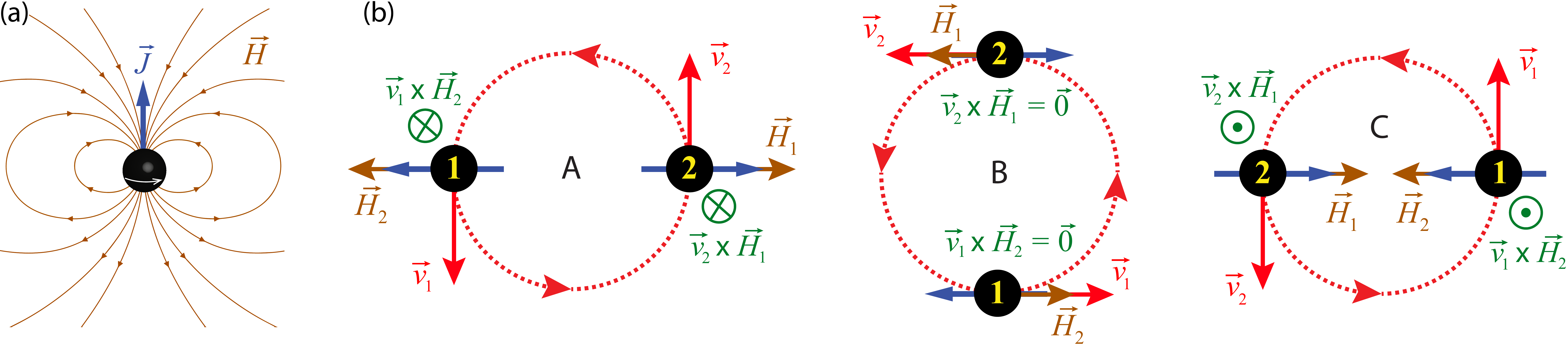}

\caption{\label{fig:TrueBobbing}(a) The field lines of the gravitomagnetic
field $\vec{H}$ produced by a spinning black hole (of spin vector
$\vec{J}$); it is similar (far from the BH) to that of a magnetic
dipole oriented oppositely to $\vec{J}$. (b) The actual picture of
the corresponding \emph{spin-orbit} gravitomagnetic forces $M\vec{v}\times\vec{H}$
involved in the bobbing motion, with phase D suppressed since the
result is similar to B. The situation is \emph{opposite} to that in
Fig. \ref{fig:FalseBobbing}: the forces vanish when the spins are
orthogonal to the axis connecting the black holes (phases B and D,
as therein $\vec{v}_{1}\parallel\vec{H}_{2}$, $\vec{v}_{2}\parallel\vec{H}_{1}$),
and have maximum magnitude when they lie along such an axis, pushing
the pair out of the plane of the illustration in phase A, and inwards
in phase C.}
\end{figure}

Frame-dragging plays a crucial role in the dynamics of binary systems.
Among other effects, it is at the origin of the bobbings observed
in numerical simulations \cite{CampanelliBobbings,Lovelace2009Bobbings}
of nearly identical black holes with anti-parallel spins along the
orbital plane (``extreme kick configurations'', see Figs. \ref{fig:FalseBobbing}-\ref{fig:TrueBobbing}),
in which the \emph{whole binary} (i.e., its center of mass) oscillates
``up and down.'' They are an example of a setting where the naive
viscous fluid analogy, and its associated ``body-dragging'' misconception,
lead to predictions opposite to the real effects.

The setting is illustrated in Fig. \ref{fig:FalseBobbing}; according
to the fluid dragging analogy, in phases A and C there would be no
dragging forces, which would be maximum in B and D, each black hole
dragging the other into the plane of the illustration in phase B,
and outwards in phase D. None of this is true, however.

The gravitomagnetic field of a body (or black hole) with spin vector
$\vec{J}$ is given, at leading order, and in its post-Newtonian (PN)
rest frame, by (e.g. \cite{CiufoliniWheeler,Costa:2015hlh}) 
\begin{equation}
\vec{H}=2\frac{\vec{J}}{r^{3}}-6\frac{(\vec{J}\cdot\vec{r})\vec{r}}{r^{5}}\ ,\label{eq:GMfieldSpinning}
\end{equation}
similar to that of a magnetic dipole, with $-2\vec{J}$ in the place
of the magnetic moment $\vec{\mu}$. Its field lines are depicted
in Fig. \ref{fig:TrueBobbing}(a). The gravitomagnetic force exerted
by black hole 1 (BH 1) on black hole 2 (BH 2) is, in the former's
PN instantaneous rest frame, $\vec{F}_{{\rm GM1,2}}=\gamma M_{2}\vec{U}_{2}\times\vec{H}_{1}\approx M_{2}\vec{v}_{21}\times\vec{H}_{1}$,
where $\vec{v}_{21}=\vec{v}_{2}-\vec{v}_{1}$ is the velocity of BH
2 relative to BH 1, and in the last equality we used $U^{\alpha}=U^{0}(1,\vec{v})$,
and approximated $\vec{U}\approx\vec{v}\equiv d\vec{x}/dt$. Using
the vector identity (5.2a) of \cite{Faye:2006gx}, it reads 
\begin{align}
\frac{\vec{F}_{{\rm GM}1,2}}{M_{2}} & =\vec{v}_{21}\times\vec{H}_{1}=-\frac{4}{r_{21}^{3}}\vec{v}_{21}\times\vec{J}_{1}-\frac{6\vec{r}_{21}[(\vec{v}_{21}\times\vec{r}_{21})\cdot\vec{J}_{1}]}{r_{21}^{5}}-\frac{6(\vec{v}_{21}\cdot\vec{r}_{21})\vec{J}_{1}\times\vec{r}_{21}}{r_{21}^{5}}\nonumber \\
 & =-\frac{4}{r_{21}^{3}}\vec{v}_{21}\times\vec{J}_{1}\ ,\label{eq:FGM12}
\end{align}
where $\vec{r}_{21}\equiv\vec{x}_{2}-\vec{x}_{1}$ is the position
vector of BH 2 relative to BH 1, and in the second equality we noticed
that $(\vec{v}_{21}\times\vec{r}_{21})\cdot\vec{J}_{1}=0$ since the
black hole's spins lie in the orbital plane, and $\vec{v}_{21}\cdot\vec{r}_{21}\approx0$
for quasi-circular motion. The analogous expression for $\vec{F}_{{\rm GM2,1}}=M_{1}\vec{v}_{12}\times\vec{H}_{2}$
follows by interchanging $1\leftrightarrow2$ in \eqref{eq:FGM12}.
It is clear from Fig. \ref{fig:TrueBobbing} that these forces vanish
in phases B and D, as therein the BHs' velocities are aligned with
their gravitomagnetic fields: $\vec{v}_{1}\parallel\vec{H}_{2}$,
$\vec{v}_{2}\parallel\vec{H}_{1}$. They have maximum magnitude in
phases A and C, when each black hole's velocity is orthogonal to the
other's spin-vector, pointing out of the plane of the illustration
in phase A, and inwards in phase C. That is, the situation is opposite
in phasing to the ``body-dragging'' picture of Fig. \ref{fig:FalseBobbing}.
(It is rather surprising that in \cite{KeppelBobbings}, where Fig.
\ref{fig:FalseBobbing} is reproduced, and at the same time equations
equivalent to the above are presented, this disagreement has seemingly
gone unnoticed).

\subsubsection{The origin of the bobbing\label{subsec:The-origin-of-bobbing}}

The gravitomagnetic forces ($\vec{F}_{{\rm GM1,2}}$ and $\vec{F}_{{\rm GM2,1}}$)
each black hole exerts on the other, which are spin-orbit interactions,
point in the same direction, seemingly causing the whole pair to accelerate.
Albeit true, it is not the whole story, as these are not the only
spin-orbit forces in this system. Each black hole also exerts a spin-curvature
force given by Eq. \eqref{eq:SpinCurvature}, which is of the same
order of magnitude {[}albeit of different nature, not an inertial
force but a ``real'', covariant force, causing the body to deviate
from geodesic motion, as discussed in Sec. \ref{subsec:Gravitomagnetic-tidal-effects:}{]},
as we shall now see.

Compact bodies such as black holes have angular momentum $J\lesssim M^{2}$,
hence the gravitomagnetic acceleration in Eq. \eqref{eq:FGM12} is
of the order $vJ/r^{3}\lesssim vM^{2}/r^{3}\sim vU(\nabla U)\sim O(5)$,
i.e., it arises at 1.5 order in the post-Newtonian (PN) expansion,
see Appendix \ref{subsec:Post-Newtonian-approximation}. Thus, we
need equations of motion for spinning (pole-dipole) bodies accurate
to 1.5 PN order. These follow by taking the 1.5 PN limit of the Mathisson-Papapetrou
Equation \eqref{eq:SpinCurvature}, which reduces (see Appendix \ref{subsec:Equations-Binaries})
to \eqref{eq:NonGeoPN}, with $\vec{F}$ the 1.5PN limit of the spin-curvature
force $F^{i}=-\mathbb{H}^{\beta i}J_{\beta}$. The force exerted by
BH 2 on BH 1 is $F_{{\rm 2,1}}^{i}=-(\mathbb{H}_{2})_{\beta}^{\ i}J_{1}^{\beta}$,
where $(\mathbb{H}_{2})_{\alpha\beta}=\star R_{\alpha\mu\beta\nu}U_{1}^{\mu}U_{1}^{\nu}$
is the gravitomagnetic tidal tensor produced by BH 2 as ``measured''
by BH 1 (Eqs. (94) and (88)-(89) in \cite{Costa:2012cy}); it reads
\cite{Costa:2012cy,Wald:1972sz,ThorneHartle1984,Magnus} 
\begin{equation}
\vec{F}_{2,1}=-\frac{3M_{2}}{r_{12}^{3}}\left[\vec{v}_{12}\times\vec{J}_{1}+\frac{2\vec{r}_{12}[(\vec{v}_{12}\times\vec{r}_{12})\cdot\vec{J}_{1}]}{r_{12}^{2}}+\frac{(\vec{v}_{12}\cdot\vec{r}_{12})\vec{J}_{1}\times\vec{r}_{12}}{r_{12}^{2}}\right]=\frac{3M_{2}}{r_{12}^{3}}\vec{v}_{21}\times\vec{J}_{1}\ ,\label{F21}
\end{equation}
where, again, in the second equality we noticed that $(\vec{v}_{12}\times\vec{r}_{12})\cdot\vec{J}_{1}=0$
for spins lying in the orbital plane, that $\vec{v}_{12}\cdot\vec{r}_{12}\approx0$
for quasi-circular motion, and that $\vec{v}_{12}=\vec{v}_{1}-\vec{v}_{2}=-\vec{v}_{21}$.
The analogous expression for $\vec{F}_{{\rm 1,2}}$ follows by interchanging
$1\leftrightarrow2$ in \eqref{F21}. Observe that it is $\vec{F}_{2,1}$
(not $\vec{F}_{{\rm GM2,1}}$) which should be regarded (in the context
of a PN approximation) as the ``reaction'' to the gravitomagnetic
force $\vec{F}_{{\rm GM1,2}}$ in Eq. \eqref{eq:FGM12}, as these
are the ones that depend on $\vec{J}_{1}$; and note that they do
not cancel out: 
\begin{equation}
\vec{F}_{{\rm GM1,2}}+\vec{F}_{2,1}=-M_{2}\frac{\vec{v}_{21}\times\vec{J}_{1}}{r_{12}^{3}}=\frac{1}{4}\vec{F}_{{\rm GM1,2}}\ne0\ .\label{eq:ActionReaction}
\end{equation}
That is, the spin-orbit interactions \emph{do not} obey an action-reaction
law \cite{Magnus}. It is this mismatch that causes the whole binary
to bob. Consider the PN frame \emph{momentarily} comoving with BH
1; the 1.5PN gravitational field generated by BH 1 is, in such frame,
described by the metric \eqref{eq:PNmetric} in Appendix \ref{subsec:Post-Newtonian-approximation},
\[
ds^{2}=(-1+2w-2U^{2})dt^{2}+2\mathcal{A}_{i}dx^{i}+\delta_{ij}\left(1+2U\right)dx^{i}dx^{j}\ ,
\]
with $w$ given by Eq. \eqref{eq:wExteriorBinary} with $\vec{v}_{1}=0$,
and $U=M_{1}/r_{1}$, $\vec{\mathcal{A}}=2\vec{r}_{1}\times\vec{J}_{1}/r_{1}^{3}$,
cf. Eqs. \eqref{eq:PNExteriorBinary}. The equation of motion for
BH 2 is then, by Eq. \eqref{eq:NonGeoPN},

\begin{equation}
\frac{d^{2}\vec{x}_{2}}{dt^{2}}=(1+v_{21}^{2}-2U)\vec{G}-3\frac{\partial U}{\partial t}\vec{v}_{21}-4(\vec{G}\cdot\vec{v}_{21})\vec{v}_{21}+\frac{\vec{F}_{{\rm GM}1,2}+\vec{F}_{1,2}}{M_{2}}+O(6)\label{eq:CoordAccelBH2}
\end{equation}
with $\vec{G}=\propto\vec{r}_{21}$ {[}cf. Eq. \eqref{eq:Gextreme}
with $\vec{v}_{1}=0${]}. The extreme kick configuration corresponds
to the case that $M_{1}=M_{2}\equiv M$, $\vec{J}_{1}=-\vec{J}_{2}$
$\in x{\rm O}y$; hence $\vec{F}_{1,2}=3M\vec{v}_{12}\times\vec{J}_{2}/r_{12}^{3}=3M\vec{v}_{21}\times\vec{J}_{1}/r_{12}^{3}=-3\vec{F}_{{\rm GM1,2}}/4$,
cf. Eq. \eqref{eq:FGM12}. Since $\vec{v}_{21}$ and $\vec{G}$ lie
both in the orbital plane $x{\rm O}y$, by \eqref{eq:CoordAccelBH2}
the $z$-coordinate acceleration of BH 2 reduces to 
\begin{equation}
\frac{d^{2}z_{2}}{dt^{2}}=\frac{F_{{\rm GM1,2}}^{z}+F_{1,2}^{z}}{M}=\frac{(\vec{J}_{1}\times\vec{v}_{21})^{z}}{r_{12}^{3}}=\frac{2vJ\cos(\Omega t)}{r_{12}^{3}}\label{eq:Bobbing}
\end{equation}
where $J=|\vec{J}_{1}|=|\vec{J}_{2}|$, $v$ is the magnitude of the
BH's velocities $\vec{v}_{2}=-\vec{v}_{1}$ with respect to the binary's
center of mass frame, $\Omega=v/r$, and in the last equality we assumed
(without loss of generality) the system to be at phase A of Fig. \ref{fig:TrueBobbing}
at $t=0$. By analogous steps one finds the same $z$-coordinate acceleration
for BH 1, $d^{2}z_{1}/dt^{2}=d^{2}z_{2}/dt^{2}$; that is, the bobbing
is synchronous, the whole binary (thus its center of mass) oscillating
up and down.

It should be stressed that such bobbing, and the mismatch between
action and reaction in the spin-orbit forces, \emph{do not} imply
a violation of any conservation principle; on the contrary, as shown
in \cite{KeppelBobbings}, it is a necessary consequence of the interchange
between mechanical momentum of the bodies and field momentum (in the
sense of the Landau-Lifshitz pseudotensor \cite{Kaplan:2009}).

It should also be remarked that in Ref. \cite{Pretorius2009}, where
the scheme in Fig. \ref{fig:FalseBobbing} was originally presented,
still the author was remarkably not misled into a qualitatively wrong
phasing. Therein a notion of ``space dragging'', seemingly different
from a viscous fluid analogy, and seemingly imparting a maximum bobbing
velocity at stages B and D is used (hence apparently closer to the
phenomena involved in the ZAMOs dragging, as if each black hole was
``trying'' to maintain constant its orbital angular momentum about
the other). Such a notion agrees in phasing with Eq. \eqref{eq:Bobbing},
which, integrating, yields $dz_{2}/dt=2vJ\sin(\Omega t)/(r_{12}^{3}\Omega)$:
it is indeed in phases B and D ($\phi=\Omega t=\pi/2$ and $\phi=3\pi/2$,
respectively) that the bobbing velocities have maximum magnitude.
However we notice that, as discussed in Sec. \ref{sec:Distinct-effects-under},
the gravitomagnetic forces (dragging of the compass of inertia) are
actually of different origin from the ZAMOs dragging; and that such
a notion is, at best, complicated and much less intuitive than the
gravitomagnetic picture.

Finally, we made use above of a PN frame momentarily comoving with
BH 1, which greatly simplified computations, since, in this case:
the gravitomagnetic field $\vec{H}_{1}$ produced by BH 1 reduces
to that generated by its spin, Eq. \eqref{eq:GMfieldSpinning}; the
gravitomagnetic force $\vec{F}_{{\rm GM1,2}}$ encodes the whole spin-orbit
inertial force acting on BH 2, and $\vec{G}$ lies along $\vec{r}_{12}$.
In other PN frames (such as one momentarily comoving with the binary's
center of mass, as depicted in Figs. \ref{fig:FalseBobbing}-\ref{fig:TrueBobbing}),
the description is more complicated, as $\vec{H}_{1}$ then includes
a contribution due to BH 1's translational motion, Eq. \eqref{eq:Hextreme},
and part of the spin-orbit inertial force is then encoded in the gravitoelectric
field $\vec{G}$ {[}having then a non-vanishing $z$-component, see
Eq. \eqref{eq:Gextreme}{]}; however, the total spin-orbit inertial
force is the same, and equal to \eqref{eq:FGM12}. Likewise, the spin-curvature
force does not depend on the chosen PN frame, so neither does the
$z$-coordinate acceleration, which holds for a generic PN frame,
cf. Eq. \eqref{eq:zaccelGen}. The derivation for a generic PN frame
is given in Appendix \ref{subsec:Equations-Binaries}.

\section{Conclusions}

Generically very feeble in the solar system, where they have been
subject of different experimental tests \cite{CiufoliniPavlisNature2004,CIUFOLINIPeron2006,LARES2019,NordtvedtPRL1988,MurphyNordtvedtTuryshev2007,SoffelKlioner2008,GPB,LucchesiLARASE2020}
(including dedicated space missions \cite{GPB,LARES2019}, as well
as some controversies \cite{KopeikinPRLComment2007,MurphyReplyPRL2007,IorioComment2017,CiufoliniComment2018}),
gravitomagnetic effects become preponderant in the strong field regime,
shaping the orbits of binary systems and the waveforms of the emitted
gravitational radiation \cite{Apostolatos:1994mx,LangHughes2006,Hannam:2013pra,Vecchio:2003tn,Schmidt:2014iyl}.
Yet they are still commonly misunderstood, in particular those dubbed
``frame-dragging''. Pertaining initially to the dragging of the
compass of inertia, the usage of the term has been extended to other
gravitomagnetic effects, as well as to persistent misconceptions fueled
by the deceptive fluid-dragging analogy --- possibly even by the
very term ``dragging''. We aimed in this paper to deconstruct such
misconceptions, explaining what the different types of frame-dragging
effects consist of, and the relationship between them. We split them
into three different levels (Table \ref{tab:Levels}), governed by
three distinct mathematical objects, corresponding to different orders
of differentiation of the gravitomagnetic potential 1-form $\bm{\mathcal{A}}$.
The first level (Sec. \ref{subsec:DraggingZAMOS}), governed by $\bm{\mathcal{A}}$
itself, may be cast physically (in axistationary spacetimes) as the
dragging of the ZAMOs, which comprises effects such as the Sagnac
effect in optical loops around the source, or the arrangement of the
bodies' angular velocities/angular momentum in black holes surrounded
by disks or rings (Sec. \ref{sec:Frame-dragging-is-never}). It contributes
also to the gravitomagnetic clock effect (Sec. \ref{subsec:Competing-effects}).
The second level, governed by the gravitomagnetic field $\vec{H}$,
and physically interpreted as the dragging of the compass of inertia
(Sec. \ref{subsec:Dragging-of-theCompass}), includes the gravitomagnetic
(or Coriolis) forces on test bodies and the Lense-Thirring orbital
precessions, gyroscope precession, and is also responsible for the
remainder of the gravitomagnetic clock effect. The third level (Sec.
\ref{subsec:Gravitomagnetic-tidal-effects:}), governed by the gravitomagnetic
tidal tensor $\mathbb{H}_{\alpha\beta}$ (``magnetic'' part of the
Riemann tensor), and interpreted as a differential dragging of the
compass of inertia, physically manifests in the relative precession
of nearby sets of gyroscopes, and in the spin-curvature force on a
gyroscope. These levels are largely \emph{independent}, in that there
exist spacetimes possessing only the first (which we exemplified with
spinning Lewis-Weyl cylinders, Sec. \ref{subsec:Competing-effects}),
and others possessing only the first two levels, but missing the third
(e.g. the Gödel universe). In the case of the first two levels (dragging
of the ZAMOs vs dragging of the compass of inertia), their effects
can actually be opposite, as exemplified here by the case of circular
geodesics in the Kerr spacetime (Sec. \ref{subsec:Competing-effects}).

Two main analogies are commonly used to help to physically interpret
the frame-dragging effects: the analogy with magnetism (which created
the term ``gravitomagnetism'') and the fluid dragging analogy. The
former is clearly useful for the second and third levels of frame-dragging,
since not only the gravitational effects comprised therein have an
electromagnetic analogue, as the \emph{exact} equations describing
them exhibit, in the appropriate formalism, exact analogies (up to
constant factors) with the electromagnetic counterparts (Secs. \ref{subsec:Dragging-of-theCompass}-\ref{subsec:Gravitomagnetic-tidal-effects:}).
As for the first level, they have no analogue in classical electromagnetism
(only in quantum electrodynamics\footnote{Namely the analogy between the Sagnac and the Aharonov-Bohm effects
(see e.g. \cite{Cilindros,RizziRuggieroAharonovII} and references
therein); it does not, however, assist much in the understanding of
the former, as the latter effect is, conceptually, more difficult.}). The fluid model, on the other hand, draws an analogy with the dragging
of a viscous fluid by a moving/rotating body. It gives some qualitative
intuition (in a loose sense) for the first level of frame-dragging,
in that in all the systems considered herein (black hole spacetimes,
and infinite spinning cylinders) the ZAMOs are dragged in the same
direction of the source's rotation. The parallelism ends there however:
other features of any fluid flow, such as the unavoidable dragging
of immersed bodies along with the flow, have no parallel in any of
the frame-dragging levels. The dragging of the compass of inertia,
in particular (which was the original motivation for such analogy),
is a very different (sometimes opposite, cf. Secs. \ref{sec:Frame-dragging-is-never}
and \ref{subsec:Bobbings-in-binary}) phenomenon from body-dragging.
Some of these inconsistencies have been pointed out in the literature,
namely in a paper by Rindler \cite{RINDLERDragging} --- based, however,
on a weak field slow motion linearized theory approach, the question
remaining as to whether the conclusions would fully hold in the exact
case. We generalized them here (Sec. \ref{sec:Frame-dragging-is-never})
to the exact theory, using the exact 1+3 GEM formalism. We considered
first (as an application akin, on the whole, to \cite{RINDLERDragging})
a space station around a spinning black hole (seen to remain stationary,
without acquiring any rotation), as well as the situation for test
particles launched from it, where those with initial outwards radial
velocity are deflected in the direction opposite to the black hole's
rotation. We considered also the reciprocal problem --- a rotating
ring around a `non-spinning' black hole (either with zero angular
momentum, or zero horizon angular velocity), pointing out that the
black hole acquires no angular acceleration of any sort, the solutions
being \emph{stationary}. As a very physical, and perhaps sharpest
example to convince the reader that the (all too common) notion of
``body-dragging'' is wrong, and its underlying viscous fluid-dragging
analogy extremely misleading, we considered equilibrium positions
for test particles in spinning black hole solutions (namely Kerr-Newman
and Kerr-de Sitter). Finally, we considered a notable phenomenon which
is driven by frame-dragging --- the bobbings in ``extreme kick''
binary systems --- and where the viscous-dragging picture predicts
the exact opposite of the real effect.

\section*{Acknowledgments}

L.F.C. and J.N. were supported by FCT/Portugal through projects UIDB/MAT/04459/2020
and UIDP/MAT/04459/2020.

\appendix

\section{Inertial forces --- general formulation\label{Appendix:GEMgeneral}}

The exact formulation of the inertial GEM fields given in Sec. \ref{subsec:Dragging-of-theCompass},
and, in particular, Eqs. \eqref{eq:QMGeo}-\eqref{eq:GEMFieldsQM},
hold for stationary fields. We briefly present here its generalization
for arbitrary fields. Several such formulations have been given in
e.g. \cite{Analogies,ManyFaces,Cattaneo1958,MassaII,BiniIntrinsic,RizziRuggieroAharonovII};
we will follow the exact approach in \cite{Analogies}, which, in
the corresponding limits, leads directly to the GEM fields usually
defined in post-Newtonian approximations, e.g.~\cite{Damour:1990pi,SoffelKlioner2008,WillPoissonBook},
and (up to constant factors and sign conventions) in the linearized
theory approximations, e.g. \cite{CiufoliniWheeler,Harris_1992,ohanian_ruffini_2013,Ruggiero:2002GMeffects,Gralla:2010xg}.

Consider a congruence of observers of 4-velocity $u^{\alpha}$, and
a (point-like) test particle of worldline $x^{\alpha}(\tau)$ and
4-velocity $dx^{\alpha}/d\tau=U^{\alpha}$. Let $U^{\langle\alpha\rangle}\equiv h_{\beta}^{\alpha}U^{\beta}$
be the spatial projection of the particle's velocity with respect
to $u^{\alpha}$, cf. Eq. \eqref{eq:SpaceProjector}; it yields, up
to a $\gamma$ ($\equiv-u_{\alpha}U^{\alpha}$) factor, the relative
velocity of the particle with respect to the observers (cf. e.g. \cite{ManyFaces,BolosIntrinsic,Costa:2012cy}).
It is the variation of $U^{\langle\alpha\rangle}$ along $x^{\alpha}(\tau)$
that one casts as \emph{inertial forces} (per unit mass); the precise
definition of such variation involves some subtleties however. For
that we need a connection $\tilde{\nabla}$ (i.e., a covariant derivative)
for spatial vectors that (i) in the directions orthogonal to $u^{\alpha}$
should equal the projected Levi-Civita spacetime connection: $\tilde{\nabla}_{\mathbf{X}}Z^{\alpha}=h_{\ \beta}^{\alpha}\nabla_{\mathbf{X}}Z^{\beta}$,
for any $X^{\alpha}$ and $Z^{\alpha}$ orthogonal to $u^{\alpha}$,
so that it corrects for the trivial variation of the spatial axes
in the directions orthogonal to $u^{\alpha}$ (for instance, of a
non-rectangular coordinate system in flat spacetime) which is not
related to inertial forces and does not vanish in an inertial frame;
(ii) along the congruence, becomes an ordinary time-derivative $\partial_{{\bf u}}$,
so that it yields the variation of $U^{\langle\alpha\rangle}$ with
respect to a system of spatial axes undergoing a transport law specific
to the chosen reference frame. The most natural of such choices is
spatial axes \emph{co-rotating} with the observers (``congruence
adapted''\emph{ }frame \cite{Analogies}; arguably, the closest generalization
of the Newtonian concept of reference frame \cite{MassaZordan,MassaII}).
For an orthonormal basis $\mathbf{e}_{\hat{\alpha}}$, whose general
transport law along the observer congruence can be written as (e.g.
\cite{Misner:1974qy,Analogies}) 
\[
\nabla_{\mathbf{u}}\mathbf{e}_{\hat{\beta}}=\Omega_{\,\,\hat{\beta}}^{\hat{\alpha}}\mathbf{e}_{\hat{\alpha}};\quad\Omega^{\alpha\beta}=2u^{[\alpha}\nabla_{\mathbf{u}}u^{\beta]}+\epsilon_{\ \ \nu\mu}^{\alpha\beta}\Omega^{\mu}u^{\nu}\ ,
\]
that amounts to choosing $\Omega^{\alpha}$ (the angular velocity
of rotation of the spatial axes relative to Fermi-Walker transport)
equal to the observer's vorticity: $\Omega^{\alpha}=\omega^{\alpha}$,
as defined in \eqref{eq:GEM Fields Cov}. If the congruence is rigid,
this ensures that the axes $\mathbf{e}_{\hat{\imath}}$ point to fixed
neighboring observers. The connection that yields the variation of
a spatial vector $X^{\alpha}$ with respect to such a frame is $\tilde{\nabla}_{\alpha}X^{\beta}\equiv h_{\gamma}^{\beta}{\nabla}_{\alpha}X^{\gamma}+u_{\alpha}\epsilon_{\ \delta\gamma\lambda}^{\beta}u^{\gamma}X^{\delta}\omega^{\lambda}$,
cf. Eq. (51) of \cite{Analogies}; and the inertial or ``gravitoelectromagnetic''
force on a test particle as measured in such frame is the variation
of $U^{\langle\alpha\rangle}$ along $x^{\alpha}(\tau)$ with respect
to $\tilde{\nabla}$, that is, $\tilde{\nabla}_{\mathbf{U}}U^{\langle\alpha\rangle}\equiv\tilde{D}U^{\langle\alpha\rangle}/d\tau$.
Since, for geodesic motion, $\nabla_{\mathbf{U}}U^{\alpha}=0$, it
follows, using \eqref{eq:SpaceProjector}, that $\tilde{\nabla}_{\mathbf{U}}U^{\langle\alpha\rangle}=-\gamma(\nabla_{\mathbf{U}}u^{\alpha}+\epsilon_{\ \delta\gamma\lambda}^{\alpha}u^{\gamma}U^{\delta}\omega^{\lambda})$,
where $\gamma\equiv-U_{\alpha}u^{\alpha}$. Finally, from the decomposition
(e.g. Eq. (135) of \cite{Invariants}) 
\begin{equation}
\nabla_{\beta}u_{\alpha}\equiv u_{\alpha;\beta}=-u_{\beta}\nabla_{{\bf u}}u_{\alpha}-\epsilon_{\alpha\beta\gamma\delta}\omega^{\gamma}u^{\delta}+\sigma_{\alpha\beta}+\frac{\theta}{3}h_{\alpha\beta}\ ,\label{eq:Kinematics-Decomp}
\end{equation}
where $\sigma_{\alpha\beta}=h_{\alpha}^{\mu}h_{\beta}^{\nu}u_{(\mu;\nu)}-\theta h_{\alpha\beta}/3$
and $\theta\equiv u_{\ ;\alpha}^{\alpha}$ are, respectively, the
congruence's shear and expansion, we have (using $\nabla_{\mathbf{U}}u^{\alpha}\equiv U^{\beta}\nabla_{\beta}u^{\alpha}$)
\cite{Analogies} 
\begin{equation}
\frac{\tilde{D}U^{\langle\alpha\rangle}}{d\tau}=\gamma\left[\gamma G^{\alpha}+\epsilon_{\ \beta\gamma\delta}^{\alpha}u^{\delta}U^{\beta}H^{\gamma}-\sigma_{\ \beta}^{\alpha}U^{\beta}-\frac{\theta}{3}h_{\beta}^{\alpha}U^{\beta}\right]\equiv\frac{F_{{\rm GEM}}^{\alpha}}{m},\label{eq:GEMforce}
\end{equation}
where the ``gravitoelectric'' $G^{\alpha}$ and ``gravitomagnetic''
$H^{\alpha}$ fields are again given by Eqs. \eqref{eq:GEM Fields Cov}
(being, respectively, minus the observers' acceleration, and twice
their vorticity), and $m$ is the particle's mass. For a congruence
of observers \eqref{eq:uLab} tangent to a time-like Killing vector
field in a stationary spacetime (or for any rigid congruence of observers
in general), $\sigma_{\alpha\beta}=0=\theta$, and so Eq. \eqref{eq:GEMforce}
reduces to Eq. \eqref{eq:QMGeo}.

\subsection{Post-Newtonian approximation\label{subsec:Post-Newtonian-approximation}}

The post-Newtonian expansion is a weak field and slow motion approximation
tailored to bound astrophysical systems. It can be cast in different,
equivalent ways; here, following \cite{Misner:1974qy,Will:1993ns,JantzenPN,Kaplan:2009,Magnus,Invariants},
we frame the expansion in terms of a small \emph{dimensionless} parameter
$\epsilon$, such that $U\sim\epsilon^{2}$, where $U$ is minus the
Newtonian potential {[}i.e., taking the Newtonian limit of Eq. \eqref{eq:StatMetric},
$\Phi=-U${]}, and the bodies' velocities are assumed such that $v\lesssim\epsilon$
(since, for bounded orbits, $v\sim\sqrt{U}$). In terms of ``forces,''
the Newtonian force $m\nabla U$ is taken to be of zeroth PN order
{[}0PN, i.e., $O(2)${]}, and each factor $\epsilon^{2}$ amounts
to a unit increase of the PN order. Time derivatives increase the
degree of smallness of a quantity by a factor $\epsilon$; for example,
$\partial U/\partial t\sim Uv\sim\epsilon U$. The 1PN expansion consists
of keeping terms up to $O(4)$ in the equations of motion~\cite{Will:1993ns}.
This amounts to retaining terms up to $O(4)$ in $g_{00}$, $O(3)$
in $g_{0i}$, and $O(2)$ in $g_{ij}$, effectively considering a
metric of the form \cite{Damour:1990pi,Kaplan:2009} 
\begin{equation}
g_{00}=-1+2w-2w^{2}+O(6);\qquad g_{i0}=\mathcal{A}_{i}+O(5);\qquad g_{ij}=\delta_{ij}\left(1+2U\right)+O(4)\label{eq:PNmetric}
\end{equation}
(actually accurate to 1.5PN), where $w$ consists of the sum of $U$
plus \emph{non-linear} terms of order $\epsilon^{4}$, $w=U+O(4)$.
For observers \eqref{eq:uLab} at rest in a given coordinate system,
by \eqref{eq:GEM Fields Cov} $G^{i}=\Gamma_{00}^{i}/g_{00}$, $H^{i}=-\epsilon_{\ k0}^{ij}\Gamma_{0j}^{k}/g_{00}$;
hence\footnote{For the expressions for the Christoffel symbols, see e.g. Eqs. (8.15)
of \cite{WillPoissonBook}, identifying $w\rightarrow U+\Psi$, $\mathcal{A}_{i}\rightarrow-4U_{i}$
in the notation therein.} 
\begin{equation}
\vec{G}=\nabla w-\frac{\partial\vec{\mathcal{A}}}{\partial t}+O(6)\ ;\quad\vec{H}=\nabla\times\vec{\mathcal{A}}+O(5)\label{eq:GEMfieldsPN}
\end{equation}
(cf. e.g. Eqs. (3.21) of \cite{Damour:1990pi}). Moreover, $\sigma_{\alpha\beta}\lesssim O(5)$,
$\theta=3\partial_{t}U+O(5)$, $U^{i}=U^{0}v^{i}=v^{i}+O(3)$, $\vec{v}\equiv d\vec{x}/dt$,
and thus Eq. \eqref{eq:GEMforce} becomes 
\begin{equation}
\frac{\vec{F}_{{\rm GEM}}}{m}\equiv\frac{\tilde{D}\vec{U}}{d\tau}=(1+v^{2})\vec{G}+\vec{v}\times\vec{H}-\frac{\partial U}{\partial t}\vec{v}+O(6)\ .\label{eq:GeoPN0}
\end{equation}
In terms of coordinate acceleration, noting that $\tilde{D}\vec{U}/d\tau=(U^{0})^{2}d^{2}\vec{x}/dt^{2}+2\vec{v}\partial_{t}U-v^{2}\vec{G}+4(\vec{G}\cdot\vec{v})\vec{v}+O(6)$,
$(U^{0})^{-2}=1-v^{2}-2U+O(4)$, we have 
\begin{equation}
\frac{d^{2}\vec{x}}{dt^{2}}=\frac{\vec{F}_{{\rm IPN}}}{m}=(1+v^{2}-2U)\vec{G}+\vec{v}\times\vec{H}-3\frac{\partial U}{\partial t}\vec{v}-4(\vec{G}\cdot\vec{v})\vec{v}+O(6)\ ,\label{eq:GeoPN}
\end{equation}
(cf. Eq. (7.17) of \cite{Damour:1990pi}), where $\vec{F}_{{\rm IPN}}$
stands for ``post-Newtonian inertial force'' (in order to distinguish
from $\vec{F}_{{\rm GEM}}$). The absence of $O(5)$ terms means that
\eqref{eq:GeoPN0}-\eqref{eq:GeoPN} are actually accurate to 1.5PN
order.

\emph{``Linear dragging''.}--- The contribution $-\partial\vec{\mathcal{A}}/\partial t$
to $\vec{G}$ in \eqref{eq:GEMfieldsPN} (and thus to $\vec{F}_{{\rm IPN}}$)
means that a time-varying gravitomagnetic vector potential induces
a velocity-independent inertial force on a test particle. For instance,
in the interior of an accelerating massive shell, inertial forces
are induced in the same direction of the shell's acceleration, as
first noted by Einstein (\cite{EinsteinMeaning}, pp. 100-102; see
also \cite{DavidsonLinearDrag,GronErikson1987}). The effect has been
studied in the framework of the exact theory in some special solutions
(e.g. \cite{LyndenBellLinearDragging2012,LyndenBellLinearDragging1998}),
and has been experimentally confirmed to high accuracy (albeit indirectly,
one may argue) in the observations of binary pulsars \cite{NordtvedtIJTP1988}.
It is sometimes dubbed ``translational dragging'' \cite{GronErikson1987}
or ``linear dragging'' \cite{LyndenBellLinearDragging2012,PfisterLinearDragging}
(of inertial frames). We note that, in spite of such denominations,
it is a component of the gravitoelectric field $\vec{G}$; as such\emph{,
in the given reference frame}, it does not directly fit into the frame-dragging
types in Table \ref{tab:Levels}. It mixes, however, with the gravitomagnetic
inertial acceleration $\vec{v}\times\vec{H}$ (dragging of the compass
of inertia) in changes of frame; for instance, as exemplified in Secs.
\ref{subsec:Bobbings-in-binary} and \ref{subsec:Extreme-kick-configuration},
an inertial force which, in a given PN frame, consists solely of the
gravitomagnetic term $m\vec{v}\times\vec{H}$, in another frame can
be partially incorporated in the term $-m\partial\vec{\mathcal{A}}/\partial t$
(this stems from the transformation laws for $\vec{G}$ and $\vec{H}$,
which exhibit a certain analogy with their electromagnetic counterparts,
see \cite{JantzenPN}).

\subsection{Non-geodesic motion}

When the test body is acted upon by a covariant force $F^{\alpha}=DP^{\alpha}/d\tau$,
and in the special case that $P^{\alpha}=mU^{\alpha}$ (no ``hidden
momentum'') with the body's rest mass $m$ constant, then Eq. \eqref{eq:GEMforce}
is readily generalized to 
\begin{equation}
m\frac{DU^{\alpha}}{d\tau}=F^{\alpha}\quad\Leftrightarrow\quad m\frac{\tilde{D}U^{\langle\alpha\rangle}}{d\tau}=F_{{\rm GEM}}^{\alpha}+F^{\alpha}\ ,\label{eq:NonGeo}
\end{equation}
and its post-Newtonian limit to 
\begin{equation}
m\frac{d^{2}\vec{x}}{dt^{2}}=\vec{F}_{{\rm IPN}}+[1+O(2)]\vec{F}+O(6)\ .\label{eq:NonGeoPN}
\end{equation}

\subsection{Equations of motion for spinning binaries\label{subsec:Equations-Binaries}}

Consider a system of isolated spinning \emph{pole-dipole} bodies interacting
gravitationally. Each body $K$ is considered under the influence
of the gravitational field ``external'' \cite{ThorneHartle1984,Damour:1990pi}
to it, described (in the harmonic gauge in \cite{WillPoissonBook,Damour:1990pi,SoffelKlioner2008,Kaplan:2009,ThorneHartle1984}),
by the metric \eqref{eq:PNmetric} with \cite{Kaplan:2009,ThorneHartle1984}
\begin{align}
w & =\sum_{A\ne K}\frac{M_{A}}{r_{A}}\left(1+2v_{A}^{2}-\sum_{B\ne A}\frac{M_{B}}{r_{AB}}-\frac{1}{2}\vec{r}_{A}\cdot\frac{d\vec{v}_{A}}{dt}-\frac{(\vec{r}_{A}\cdot\vec{v}_{A})^{2}}{2r_{A}^{2}}\right)+2\sum_{A\ne K}\frac{(\vec{v}_{A}\times\vec{J}_{A})\cdot\vec{r}_{A}}{r_{A}^{3}}\ ;\label{eq:w_Nbodies}\\
\vec{\mathcal{A}} & =-4\sum_{A\ne K}\frac{M_{A}}{r_{A}}\vec{v}_{A}-2\sum_{A\ne K}\frac{\vec{J}_{A}\times\vec{r}_{A}}{r_{A}^{3}}\ ;\qquad U=\sum_{A\ne K}\frac{M_{A}}{r_{A}}\ ,\label{eq:PNPot_Nbodies}
\end{align}
accurate to 1.5PN order. Here $\vec{r}_{AB}\equiv\vec{x}_{A}-\vec{x}_{B}$,
$\vec{r}_{A}\equiv\vec{x}-\vec{x}_{A}$, $\vec{x}$ is the point of
observation, $\vec{x}_{A}$ is the instantaneous position of body
``$A$'', and $\vec{v}_{A}=d\vec{x}_{A}/dt=-d\vec{r}_{A}/dt$ its
velocity. To this accuracy, $d\vec{v}_{A}/dt$ is to be taken \emph{in
expression \eqref{eq:w_Nbodies}} as the Newtonian ``acceleration''
caused by the other bodies, $d\vec{v}_{A}/dt=-\sum_{B\ne A}M_{B}\vec{r}_{AB}/r_{AB}^{3}+O(4)$.

For a binary of compact bodies (1 and 2), the metric ``seen'' by
body 2 ($K=2$) reduces to \eqref{eq:PNmetric} with 
\begin{align}
w & =\frac{M_{1}}{r_{1}}\left(1+2v_{1}^{2}-\frac{M_{2}}{r_{12}}+\frac{1}{2}\vec{r}_{1}\cdot\vec{r}_{12}\frac{M_{2}}{r_{12}^{3}}-\frac{(\vec{r}_{1}\cdot\vec{v}_{1})^{2}}{2r_{1}^{2}}\right)+2\frac{(\vec{v}_{1}\times\vec{J}_{1})\cdot\vec{r}_{1}}{r_{1}^{3}}\ ;\label{eq:wExteriorBinary}\\
\vec{\mathcal{A}} & =-4\frac{M_{{\rm 1}}}{r_{1}}\vec{v}_{1}-2\frac{\vec{J}_{1}\times\vec{r}_{1}}{r_{1}^{3}}\ ;\qquad U=\frac{M_{{\rm 1}}}{r_{1}}\ .\label{eq:PNExteriorBinary}
\end{align}
The equation of motion for body 2 is the 1.5 PN limit of the Mathisson-Papapetrou
Equation \eqref{eq:SpinCurvature}. Under the Mathisson-Pirani spin
condition employed in \eqref{eq:SpinCurvature}, $dM_{2}/d\tau=0$
and, to the accuracy at hand, one can take $P_{2}^{\alpha}\approx M_{2}U_{2}^{\alpha}$
(see \cite{Costa:2017kdr}), hence $DP_{2}^{\alpha}/d\tau\approx M_{2}DU_{2}^{\alpha}/d\tau$
(consistent also, to the accuracy at hand, with the Tulczyjew-Dixon
and Ohashi-Kyrian-Semerák spin conditions; see \cite{Costa:2017kdr,Costa:2014nta,KyrianSemerak}
and Sec. 3.1 of the Supplement in \cite{Costa:2012cy}), leading,
by Eq. \eqref{eq:NonGeoPN}, to 
\begin{equation}
M_{2}\frac{d^{2}\vec{x}_{2}}{dt^{2}}=\vec{F}_{{\rm IPN}}+\vec{F}_{1,2}+O(6)\ ,\label{eq:accelcoor1}
\end{equation}
with the inertial force $\vec{F}_{{\rm IPN}}$ given by Eqs. \eqref{eq:GeoPN},
\eqref{eq:GEMfieldsPN}, and \eqref{eq:wExteriorBinary}-\eqref{eq:PNExteriorBinary},
and $\vec{F}_{1,2}$ the 1.5PN limit of the spin-curvature force exerted
by body 1 on body 2, $F_{{\rm 1,2}}^{i}=-(\mathbb{H}_{1})_{\beta}^{\ i}J_{2}^{\beta}$,
where $(\mathbb{H}_{1})_{\alpha\beta}=\star R_{\alpha\mu\beta\nu}U_{2}^{\mu}U_{2}^{\nu}$
is the gravitomagnetic tidal tensor produced by body 1 as ``measured''
by body 2 (Eqs. (94) and (88)-(89) in \cite{Costa:2012cy}). The latter
force reads 
\begin{equation}
\vec{F}_{1,2}=-\frac{3M_{1}}{r_{12}^{3}}\left[\vec{v}_{21}\times\vec{J}_{2}+\frac{2\vec{r}_{21}[(\vec{v}_{21}\times\vec{r}_{21})\cdot\vec{J}_{2}]}{r_{12}^{2}}+\frac{(\vec{v}_{21}\cdot\vec{r}_{21})\vec{J}_{2}\times\vec{r}_{21}}{r_{12}^{2}}\right]\label{F12}
\end{equation}
(cf. \cite{Costa:2012cy,Wald:1972sz,ThorneHartle1984,Magnus}), where
$\vec{r}_{21}\equiv\vec{x}_{2}-\vec{x}_{1}$, and $\vec{v}_{21}=\vec{v}_{2}-\vec{v}_{1}$.

\subsubsection{Extreme kick configuration\label{subsec:Extreme-kick-configuration}}

In this configuration the bodies are assumed to be two black holes
with approximately equal masses $M_{1}\approx M_{2}$, and spins equal
in magnitude but anti-parallel, $\vec{J}_{1}\approx-\vec{J}_{2}$,
lying in the orbital plane $x{\rm O}y$. They are also assumed in
a nearly circular orbit about the binary's center of mass. In this
case, $(\vec{v}_{21}\times\vec{r}_{21})\cdot\vec{J}_{2}=0$ and $\vec{v}_{21}\cdot\vec{r}_{21}\approx0$
(similar relations holding for $\vec{v}_{1}$ or $\vec{v}_{2}$ in
the place of $\vec{v}_{21}$, and $\vec{J}_{1}$) and so Eq. \eqref{F12}
reduces to $\vec{F}_{1,2}=-3M_{1}\vec{v}_{21}\times\vec{J}_{2}/r_{12}^{3}$.
By Eqs. \eqref{eq:GEMfieldsPN}, \eqref{eq:wExteriorBinary}-\eqref{eq:PNExteriorBinary},
the gravitoelectric and gravitomagnetic fields as ``seen'' by BH
2 are 
\begin{align}
\vec{G} & =-M_{1}\frac{\vec{r}_{21}}{r_{12}^{3}}\left(1-2\frac{M_{1}}{r_{21}}+2v_{1}^{2}\right)+5\frac{M_{1}M_{2}}{r_{12}^{4}}\vec{r}_{21}+\frac{4\vec{v}_{1}\times\vec{J}_{1}}{r_{21}^{3}}\label{eq:Gextreme}\\
\vec{H}_{1} & =\vec{H}_{{\rm spin}}+\vec{H}_{{\rm trans}};\qquad\vec{H}_{{\rm spin}}=2\frac{\vec{J}_{1}}{r_{12}^{3}}-6\frac{(\vec{J}_{1}\cdot\vec{r}_{21})\vec{r}_{21}}{r_{12}^{5}};\qquad\vec{H}_{{\rm trans}}=-4\frac{M_{1}}{r_{12}^{3}}\vec{v}_{1}\times\vec{r}_{21}\label{eq:Hextreme}
\end{align}
where we noticed that, at BH 2's position ($\vec{x}=\vec{x}_{2}$),
$\vec{r}_{1}\equiv\vec{x}-\vec{x}_{1}=\vec{r}_{21}$. The last term
of $\vec{G}$ tells us that, in a generic reference frame, due to
the translational motion of the source (BH 1), part of the spin-orbit
\emph{inertial} force is incorporated in the gravitoelectric field.
Equation \eqref{eq:Hextreme} tells us that, in addition to the term
$\vec{H}_{{\rm spin}}$ due to BH 1's spin, there is also the translational
contribution $\vec{H}_{{\rm trans}}$ to the gravitomagnetic field
$\vec{H}_{1}$ produced by BH 1. The gravitomagnetic force exerted
on BH 2 reads 
\begin{equation}
\vec{F}_{{\rm GM1,2}}=M_{2}\vec{v}_{2}\times\vec{H}_{1}=-\frac{4M_{2}}{r_{21}^{3}}(\vec{v}_{2}\times\vec{J}_{1}+M_{1}v_{1}v_{2}\vec{r}_{21})\ ,\label{eq:FGM12Appendix}
\end{equation}
where in $\vec{v}_{2}\times\vec{H}_{{\rm trans}}$ we again used the
vector identity (5.2a) of \cite{Faye:2006gx}. Since $\vec{v}_{1}\parallel\vec{v}_{2}\perp\vec{r}_{21}$,
the last two terms of $\vec{F}_{{\rm IPN}}$ in \eqref{eq:GeoPN}
both vanish at BH 2's position: $\partial_{t}U=M_{1}(\vec{r}_{21}\cdot\vec{v}_{1})/r_{12}^{3}=0$;
$(\vec{G}\cdot\vec{v}_{2})\vec{v}_{2}=-M_{1}(\vec{r}_{21}\cdot\vec{v}_{2})\vec{v}_{2}/r_{12}^{3}+O(6)=0.$
The equation of motion for BH 2 is thus, by \eqref{eq:accelcoor1},
\begin{align}
\frac{d^{2}\vec{x_{2}}}{dt^{2}} & =\frac{\vec{F}_{{\rm IPN}}+\vec{F}_{{\rm 1,2}}}{M_{2}}=(1+v_{2}^{2}-2\frac{M_{{\rm 1}}}{r_{1}})\vec{G}+\frac{\vec{F}_{{\rm GM1,2}}+\vec{F}_{{\rm 1,2}}}{M_{2}}\ ,\label{eq:accelcoor2}
\end{align}
with $\vec{G}$, $\vec{F}_{{\rm GM1,2}}$ and $\vec{F}_{{\rm 1,2}}$
given by Eqs. \eqref{eq:Gextreme}, \eqref{eq:FGM12Appendix}, and
\eqref{F12}. Splitting the PN inertial force into ``monopole''
and spin-orbit parts, $\vec{F}_{{\rm IPN}}=\vec{F}_{{\rm mono}}+\vec{F}_{{\rm ISO}}$,
we can re-write 
\begin{align}
 & \frac{d^{2}\vec{x}_{2}}{dt^{2}}=\frac{\vec{F}_{{\rm mono}}+\vec{F}_{{\rm ISO}}+\vec{F}_{1,2}}{M_{2}}\ ;\qquad\vec{F}_{{\rm ISO}}=-\frac{4M_{2}\vec{v}_{21}\times\vec{J}_{1}}{r_{21}^{3}}\ ;\qquad\vec{F}_{1,2}=-\frac{3M_{1}\vec{v}_{21}\times\vec{J}_{2}}{r_{12}^{3}}\ ;\label{eq:accelcoor3}\\
 & \vec{F}_{{\rm mono}}=-\frac{M_{1}M_{2}}{r_{12}^{3}}\left(1-\frac{4M_{1}+5M_{2}}{r_{12}}+v_{2}^{2}+2v_{1}^{2}+4v_{1}v_{2}\right)\vec{r}_{21}\ .\nonumber 
\end{align}
Hence $\vec{F}_{{\rm ISO}}$ matches the gravitomagnetic force \eqref{eq:FGM12}
exerted on BH 2 as measured in the PN frame momentarily comoving with
BH 1, obtained in Sec. \ref{subsec:Bobbings-in-binary}. That is:
in a generic frame, the gravitomagnetic force does not encode the
whole spin-orbit inertial force, part of it being encoded in the gravitoelectric
force $M_{2}\vec{G}$; however, the \emph{total} spin-orbit inertial
force is the \emph{same in all PN frames}. Since $\vec{F}_{{\rm mono}}\ \in\ xOy$,
we have, for the $z$- coordinate, 
\begin{equation}
\frac{d^{2}z}{dt^{2}}=\frac{\vec{F}_{{\rm ISO}}+\vec{F}_{1,2}}{M}=-\frac{(\vec{v}_{21}\times\vec{J}_{1})^{z}}{r_{12}^{3}}\ ,\label{eq:zaccelGen}
\end{equation}
matching the result \eqref{eq:Bobbing} obtained in Sec. \ref{subsec:Bobbings-in-binary}.

\bibliographystyle{utphys} {\footnotesize{}\bibliography{Ref}
}{\footnotesize\par}
\end{document}